\documentclass[iop,twocolumn,apj]{emulateapj}
\usepackage{apjfonts}
\epsscale{1.15}

\shorttitle{Radio-Selected Binary AGNs} 
\shortauthors{Fu et al.}
\journalinfo{}
\submitted{}

\newcommand{\kms}{{km s$^{-1}$}}
\newcommand{\msun}{$M_{\odot}$}

\newcommand{\um}{$\mu$m}
\newcommand{\uJy}{$\mu$Jy}

\newcommand{\ergs}{erg~s$^{-1}$}
\newcommand{\OIII}{[O\,{\sc iii}]}

\newcommand{\nod}{\nodata}
\newcommand{\NOD}{\nodata}

\begin{document}

\title{Radio-Selected Binary Active Galactic Nuclei from the Very Large Array Stripe 82 Survey\altaffilmark{*}}

\altaffiltext{*}{Some of the data presented herein were obtained at the W.M. Keck Observatory, which is operated as a scientific partnership among the California Institute of Technology, the University of California and the National Aeronautics and Space Administration. The Observatory was made possible by the generous financial support of the W.M. Keck Foundation.}

\author{
Hai~Fu\altaffilmark{1}, A.~D.~Myers\altaffilmark{2}, S.~G.~Djorgovski\altaffilmark{3}, Lin~Yan\altaffilmark{3}, J.~M.~Wrobel\altaffilmark{4}, A.~Stockton\altaffilmark{5} 
}
\altaffiltext{1}{Department of Physics \& Astronomy, University of Iowa, Iowa City, IA 52245}
\altaffiltext{2}{Department of Physics \& Astronomy, University of Wyoming, Laramie, WY 82071}
\altaffiltext{3}{California Institute of Technology, 1200 E. California Blvd., Pasadena, CA 91125}
\altaffiltext{4}{National Radio Astronomy Observatory, P.O. Box O, Socorro, NM 87801}
\altaffiltext{5}{Institute for Astronomy, University of Hawaii, 2680 Woodlawn Drive, Honolulu, HI 96822}

\begin{abstract} 
Galaxy mergers play an important role in the growth of galaxies and their supermassive black holes. Simulations suggest that tidal interactions could enhance black hole accretion, which can be tested by the fraction of binary active galactic nuclei (AGNs) among galaxy mergers. But determining the fraction requires a statistical sample of binaries. We have identified kpc-scale binary AGNs directly from high-resolution radio imaging. Inside the 92 deg$^2$ covered by the high-resolution Very Large Array survey of the Sloan Digital Sky Survey (SDSS) Stripe 82 field, we identified 22 grade A and 30 grade B candidates of binary radio AGNs with angular separations less than 5\arcsec\ (10 kpc at $z = 0.1$). Eight of the candidates have optical spectra for both components from the SDSS spectroscopic surveys and our Keck program. Two grade B candidates are projected pairs, but the remaining six candidates are all compelling cases of binary AGNs based on either emission line ratios or the excess in radio power compared to the H$\alpha$-traced star formation rate. Only two of the six binaries were previously discovered by an optical spectroscopic search. Based on these results, we estimate that $\sim$60\% of our binary candidates would be confirmed once we obtain complete spectroscopic information. We conclude that wide-area high-resolution radio surveys offer an efficient method to identify large samples of binary AGNs. These radio-selected binary AGNs complement binaries identified at other wavelengths and are useful for understanding the triggering mechanisms of black hole accretion.
\end{abstract}

\keywords{galaxies: active --- galaxies: interactions --- galaxies: nuclei --- radio continuum: galaxies}

\section{Introduction} \label{sec:introduction}

As it became increasingly clear that most galactic nuclei should host supermassive black holes \citep[e.g.][]{Kormendy95}, it was expected that binary supermassive black holes (SMBHs) should be discovered as a common byproduct of merging galaxies and that they should produce distinct signatures as a function of wavelength and separation \citep[e.g.][]{Begelman80}. Arguments based on hydrodynamical N-body simulations, coupled with relativistic considerations, suggest that
merging galaxies could trigger a range of outcomes for SMBH pairs \citep[e.g.][]{Milosavljevic01,Milosavljevic03,Escala04,Madau04}. Consequently, the exact demise of the majority of SMBH pairs remains a complex problem with few strong observational
constraints \citep[e.g.][]{Berczik06,Bonning07,Dotti07,Baker08,Berentzen09}.

Identifying inactive binary SMBHs separated by a few parsecs is a difficult observational task.
A less taxing approach is to detect pairs of Active Galactic Nuclei (AGN) on kiloparsec scales,
so-called ``dual'' or ``binary'' AGN. As dynamical models of active black holes in merging galaxies become more sophisticated, observations of binary AGN are becoming critical in order to establish the starting conditions for such models \citep[see, e.g.][]{Van-Wassenhove12,Blecha12,Blecha13,Colpi14}. 
Models that follow binary AGN in merging galaxies below kiloparsec scales help to 
set the boundary conditions for models of binary SMBH coalescence, which in turn
inform the instrumental characteristics for gravitational wave experiments 
\citep[e.g.][]{Peters64,Scheel09,Dotti12,Sesana13} such as PTAs 
\citep[Pulsar Timing Arrays, e.g.][]{Hobbs12,Tanaka13,Ravi14}.

Quite a few kpc-scale binary AGNs have now been discovered using a range of multi-wavelength
imaging techniques \citep[e.g.][]{Junkkarinen01,Komossa03,Bianchi08,Koss11,Fabbiano11}. In particular, a large number of moderate-resolution optical spectra of $z~\leq~1$ AGNs were surveyed by the Sloan Digital Sky Survey (SDSS; \citealt{York00,Schneider10})
and DEEP2 Galaxy Redshift Survey \citep{Newman13}. This triggered a boom in the study of pairs of SMBHs by prompting 
compilations of AGN spectra with double-peaked \OIII\ emission \citep{Gerke07,Comerford09a,Wang09,Smith10,Liu10a}. A small fraction ($\sim$2\%) of these spectra may represent Doppler-shifted emission lines from pairs of AGNs with transverse separations of $\sim$0.1--10 kpc \citep[e.g.][]{Shen11,Fu12a}. But, for the majority of these sources, care must be taken to distinguish genuine kpc-scale dual AGN from other phenomena such as gas outflows. 

The heterogeneous techniques that have been used to detect binary AGN, coupled with the
alternate phenomena that can describe double-peaked spectral emission lines, implies that
significant progress in this field will require a statistical sample of binary AGNs with well-understood  selection biases. The most direct way to search for binary AGNs is to use galaxy mergers identified with optical or near-infrared (IR) images. But, the vast majority of such mergers are expected to contain either inactive or single AGNs. So, the addition of high-resolution radio or X-ray imaging, or spatially resolved spectroscopy, 
is necessary to truly identify and characterize two active components in such mergers.

This work is motived by the discovery of SDSS J150243.1$+$111557, a double-peaked [O\,{\sc iii}] AGN at $z = 0.39$ that we confirmed as a binary AGN with high-resolution multi-frequency radio images from the Very Large Array \citep[VLA;][]{Fu11b}, the European VLBI Network \citep[EVN;][]{Deane14}, and the Very Long Baseline Array \citep[VLBA;][]{Wrobel14}. In fact, radio emission is known to trace AGN activity since the first AGNs were discovered \citep[i.e., Quasars;][]{Sandage61,Schmidt63}. Moreover, one of the first binary AGNs was discovered because of its double radio cores and jets --- 3C~75 at the center of Abell 400 \citep{Owen85,Beers92,Hudson06},
and the first parsec-scale binary SMBH (0402$+$379) was discovered in the radio \citep{Maness04,Rodriguez06}.
So one could search for binary AGNs directly with radio images. But the rarity of binary AGNs requires that we have a large sky coverage. Fortunately, high-resolution radio maps have already been obtained by the VLA for a large patch of the sky. Here we conducted a systematic search of binary AGNs based on the wide-area high-resolution VLA survey of the SDSS Stripe 82 field \citep{Hodge11}.

Throughout we adopt the AB magnitude system and the {\it WMAP} 7-year concordance $\Lambda$CDM cosmology with $\Omega_{\rm b}=0.0455$, $\Omega_{\rm m}=0.27$, $\Omega_\Lambda=0.73$, $H_0$ = 70~km~s$^{-1}$~Mpc$^{-1}$ \citep{Komatsu11}.

\section{Imaging Data and Pair Selection}

\begin{figure}[!t]
\plottwo{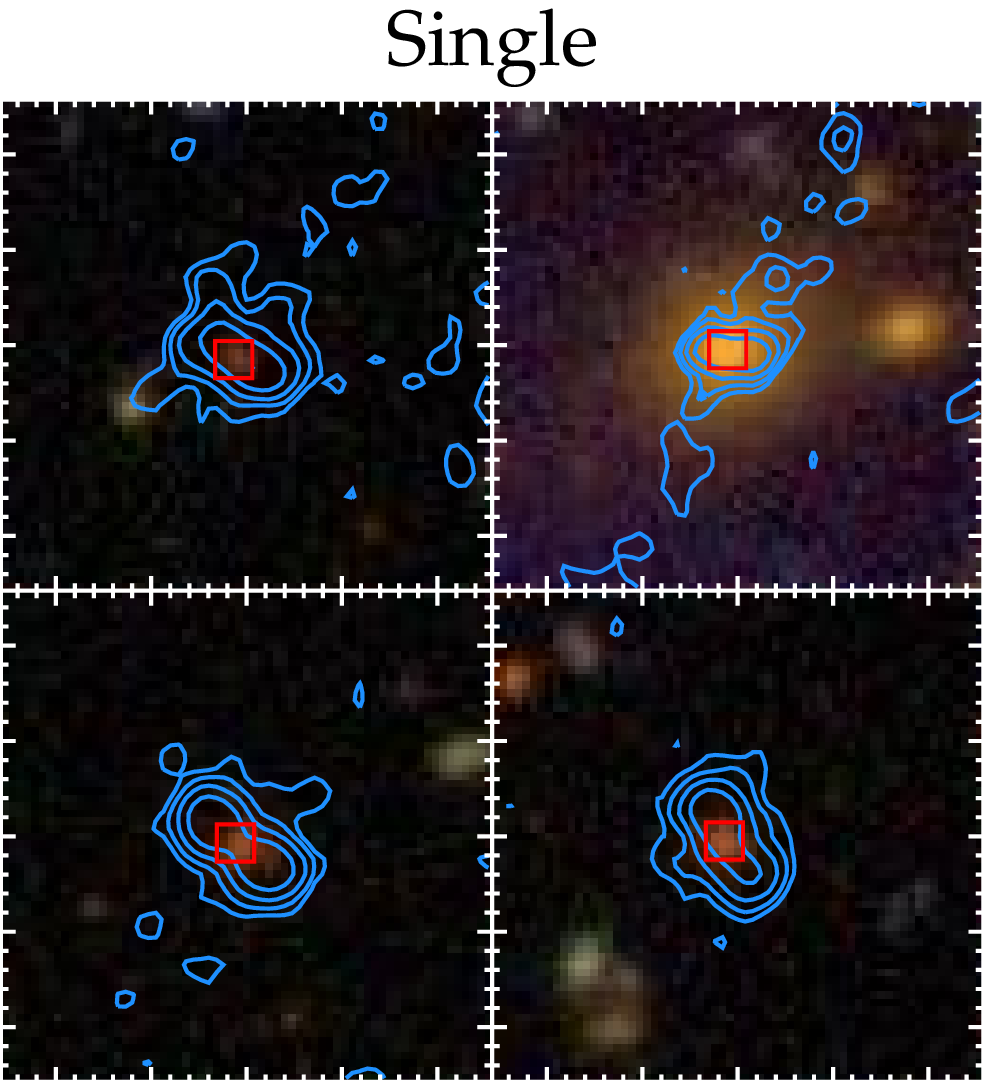}{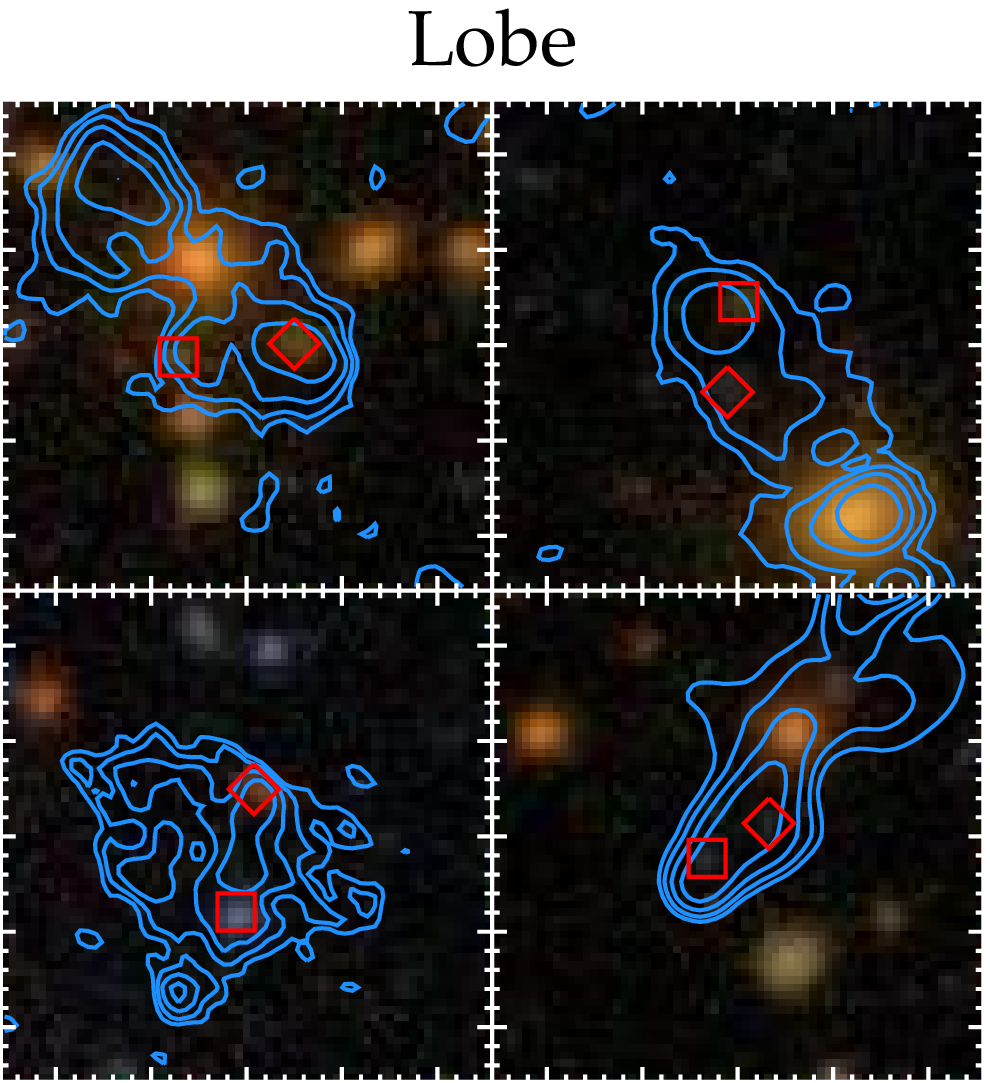}
\caption{Examples of contaminants: ($a$) singles, and ($b$) lobes. Background are SDSS $gri$ color images generated from the depth-optimized co-adds \citep{Jiang14}. Blue contours are 1.4~GHz images from the VLA-Stripe82 survey \citep{Hodge11}. The lowest contour is at 2$\sigma$ and the levels increase exponentially to the peak value of each map. Red squares and diamonds indicate the optical counterparts of the radio sources. All images are 26\arcsec\ on a side, and the major tickmarks are spaced in 5\arcsec\ intervals. N is up and E is to the left.
\label{fig:single}}
\end{figure}

\begin{figure}[!t]
\plottwo{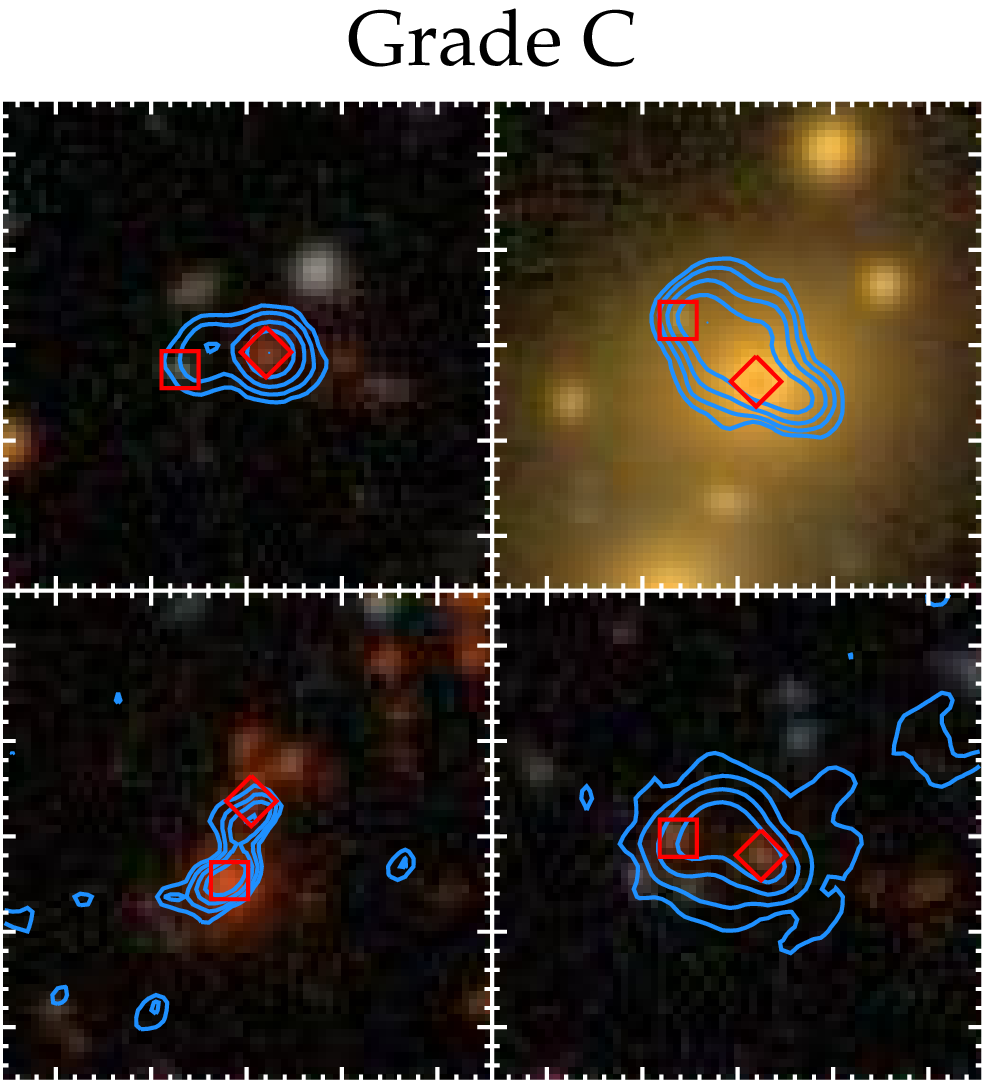}{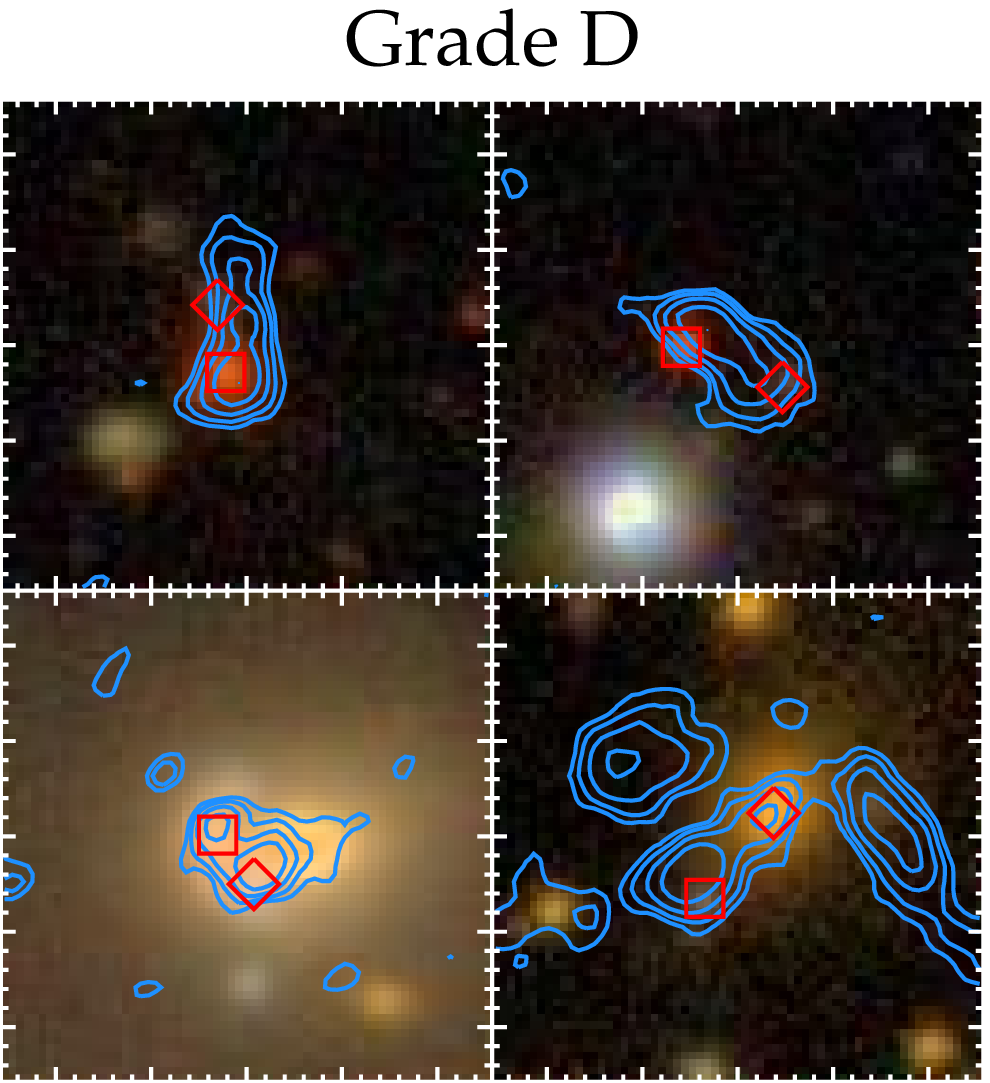}
\caption{Same as Fig.~\ref{fig:single} but for examples of misidentified IDs: Grade C and Grade D pairs.
\label{fig:gradeCD}}
\end{figure}

\begin{figure*}[!t]
\plotone{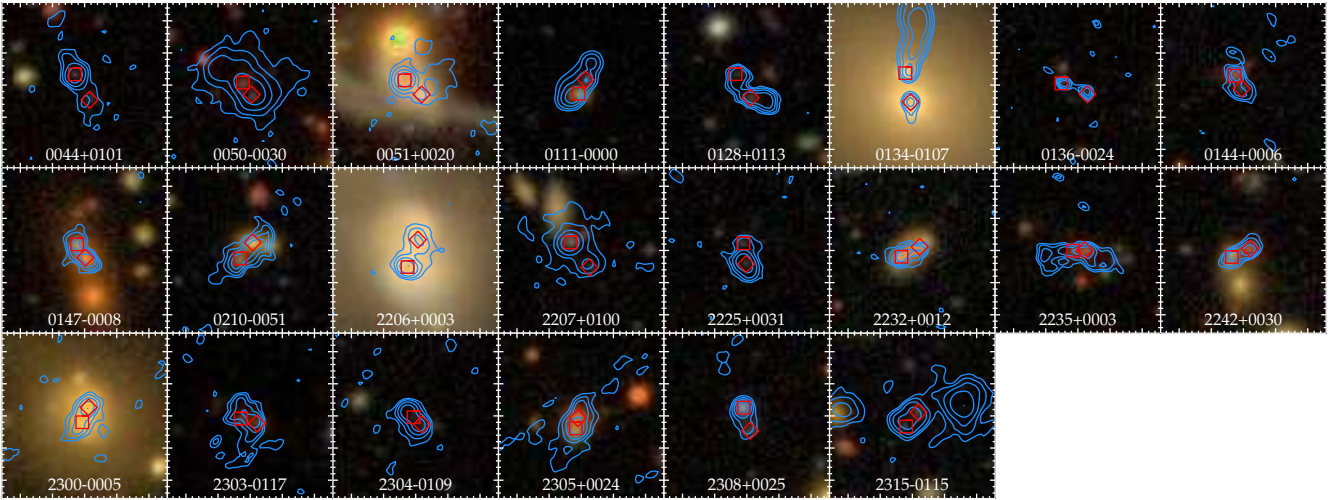}
\caption{Same as Fig.~\ref{fig:single} but for all Grade A candidate binary radio AGNs. 
\label{fig:gradeA}}
\end{figure*}

\begin{figure*}[!t]
\plotone{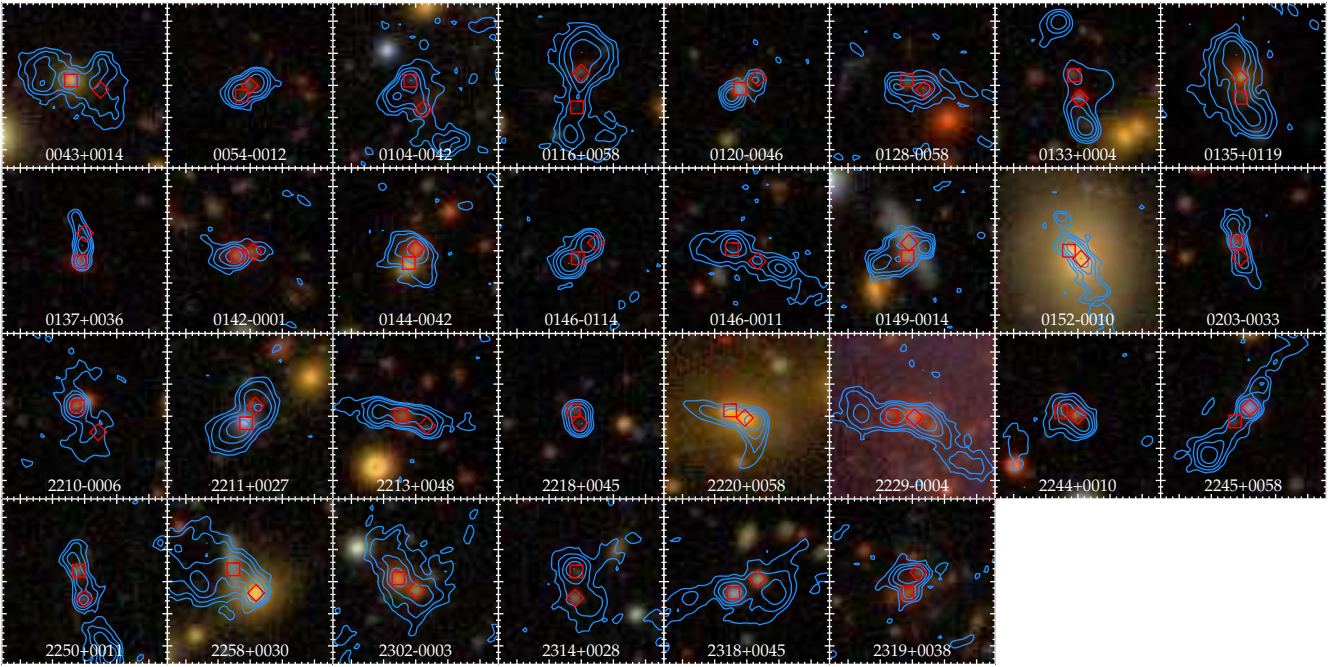}
\caption{Same as Fig.~\ref{fig:single} but for all Grade B candidates.
\label{fig:gradeB}}
\end{figure*}

The Sloan Digital Sky Survey (SDSS) Stripe 82 area is a narrow equatorial stripe in the south Galactic cap. It covers a continuous $\sim$300 deg$^2$ field centered on the vernal equinox with 20 hr $<$ R.A. $<$ 4 hr and $-1.26^\circ <$ Dec $< +1.26^\circ$. This area has been repeatedly scanned 70$-$90 times by the SDSS imaging survey, resulting in $\sim$2 mag deeper data than the best SDSS single-epoch data. \citet{Jiang14} released the latest depth-optimized co-added SDSS $ugriz$ images and SExtractor \citep{Bertin96} source catalogs of Stripe 82. The average 5$\sigma$ point-source detection limits of the co-added $ugriz$ images are respectively 23.9, 25.1, 24.6, 24.1, and 22.8 AB, and the average point-spread function (PSF) FWHM is $\sim$1\arcsec. These co-adds show significant improvement in depth and image quality over the earlier co-adds of \citet{Annis11}.

About a third of the Stripe 82 area has been mapped by the VLA at 1.4~GHz in its most extended configurations between 2007 and 2009 \citep{Hodge11}. The VLA-Stripe82 survey covers a total area of $\sim$92~deg$^2$ over two separate fields within Stripe 82: (1) 22.1 hr $<$ R.A. $<$ 23.3 hr and $-1.2^\circ <$ Dec $< +1.2^\circ$, and (2) 0.6 hr $<$ R.A. $<$ 2.4 hr and $-1.2^\circ <$ Dec $< +1.2^\circ$. The VLA survey has an angular resolution of 1.8\arcsec\ and a median rms noise of 52~\uJy~beam$^{-1}$. In comparison, the FIRST survey \citep{Becker95} has a typical resolution of 5\arcsec\ and a median rms noise of 150~\uJy~beam$^{-1}$. The improved depth and resolution of the VLA-Stripe82 survey make it ideal for searches of binary radio AGNs. 

We begin by identifying the optical counterparts of the 17,969 discrete radio sources in the VLA-Stripe82 catalog. We use both the co-added catalogs of \citet{Jiang14} and the SDSS DR10 photometric catalog \citep{Ahn14}. Considering (1) the resolution of the radio image, (2) the relative uncertainty in the astrometry between radio and optical data, and (3) source deblending errors due to complex radio morphologies, we decide to adopt a conservative matching radius of 1.8\arcsec, which equals the FWHM of the synthesized beam of the radio map. The catalogs of the co-added SDSS images are not band-matched, so each band has different source lists. To include cases where the optical counterparts in a pair have very different colors, we consider a radio source to have an optical ID if an optical source is found within the matching radius in {\it any} of the five SDSS bands. We avoid cases where multiple radio sources share the same optical counterpart by associating the optical source with the closest radio source only. We identified optical counterparts for 11,203 (62\%) radio sources. 

We then search for radio pairs with angular separations less than 5\arcsec. We choose such a small separation because (1) we are mainly interested in kpc-scale binaries (5\arcsec\ = 10 kpc at $z=0.11$) and (2) to reduce contamination from projected pairs (since we only have limited spectroscopic redshifts for the radio sources, see \S~\ref{sec:spec}). We also require both radio sources in a pair to have optical IDs to reduce contamination by extended radio sources such as multiple hotspots/jets emergent from the same radio galaxy. Considering the surface density of radio sources with optical counterparts is $\sim$122 deg$^{-2}$ in the VLA-Stripe82 survey, the probability that two random radio sources are within 5\arcsec\ is only 0.074\%. Over the entire area of the survey, we would therefore expect to find only $\sim$8 pairs by chance. In contrast, we find 134 such pairs. Because some sources belong to multiple pairs, these pairs contain 261 radio sources. Note that the separations between the optical IDs may be greater than 5\arcsec\ because of the adopted 1.8\arcsec\ matching radius.

Next, we examine the optical and radio images of these pairs to remove the following clear contaminants (see Fig.~\ref{fig:single} for examples): 
\begin{description}
\item[{\bf Singles}] A single optical source is assigned to two separate radio sources due to cross-band astrometry errors; this problem is particularly significant for faint sources. Or, one of the optical IDs is a spurious source, e.g., because of nearby bright stars, over-deblending, or edge effects (35 cases).
\item[{\bf Lobes}] Optical sources projected in extended radio lobes of a nearby radio galaxy. Based on the mean source density in the SDSS co-adds ($\sim$16 arcmin$^{-2}$ at $r < 24.6$), we expect a background density of $\sim$6 arcmin$^{-2}$ for random optical pairs with separations less than 5\arcsec\ (16 cases).
\end{description}

Finally, the remaining 83 pairs are divided into the following four grades. This step is necessary because we have adopted a relatively large matching radius between radio and optical sources (1.8\arcsec) when compared with the pair separation ($\leq$5\arcsec). 
\begin{description}
\item[{\bf Grade A}] Both optical sources are well enclosed by the radio structure and are clearly associated with a discrete radio source (i.e., ``secure optical IDs''). In such cases, the PA between the optical IDs is well aligned with that of the radio structure (i.e., ``aligned PAs'') (22 cases).
\item[{\bf Grade B}] Secure optical ID and aligned PAs, but the radio morphologies suggest that one of the optical components could be either a projected source within the radio lobe generated by the primary component or emission-line gas ejected by the radio outflow. Projected pairs are a significant concern because of the high source density of the deep co-added SDSS images. On average, $\sim$36\% of optical sources in the SDSS co-adds would have a nearby optical source within 5\arcsec.  However, we note that there is no clear boundary between grade A and B and the separation is quite subjective (30 cases).
\item[{\bf Grade C}] One of the optical IDs shows significant offset from the corresponding radio sources. But the optical and radio PAs are aligned (13 cases).
\item[{\bf Grade D}] Offset optical IDs as in grade C, but the PAs are misaligned (18 cases).
\end{description}

We consider grade C and D sources as misidentified optical IDs (see Fig.~\ref{fig:gradeCD} for examples), and grade A and B sources as candidate binary radio AGNs (see Figs.~\ref{fig:gradeA} and \ref{fig:gradeB}). The properties of these candidates are tabulated in Tables~\ref{tab:gradeA} and \ref{tab:gradeB}. 

In summary, using the VLA-Stripe82 survey and the depth-optimized SDSS co-adds, we could place a strict upper limit on the surface density of binary radio AGNs with 1.8\arcsec\ $<$ Sep $<$ 5.0\arcsec\ of $<$0.6~deg$^{-2}$. Although excluded from this study, grade C and D pairs are also interesting because one of the galaxies could be experiencing harassment from the radio outflow of the neighboring galaxy, if they are at the same redshift. 

\section{Optical Spectroscopy} \label{sec:spec}

\begin{figure*}[!t]
\plottwo{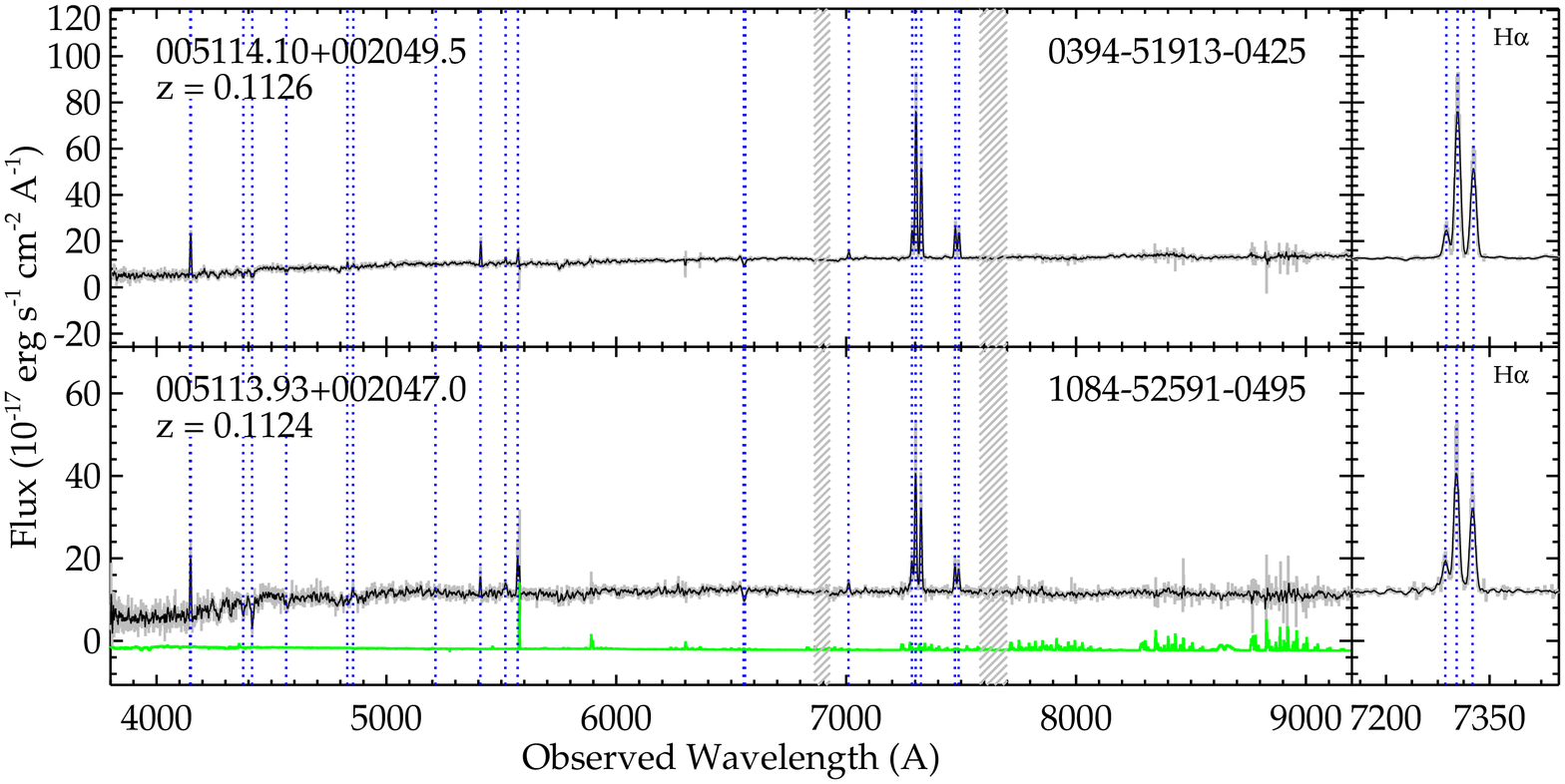}{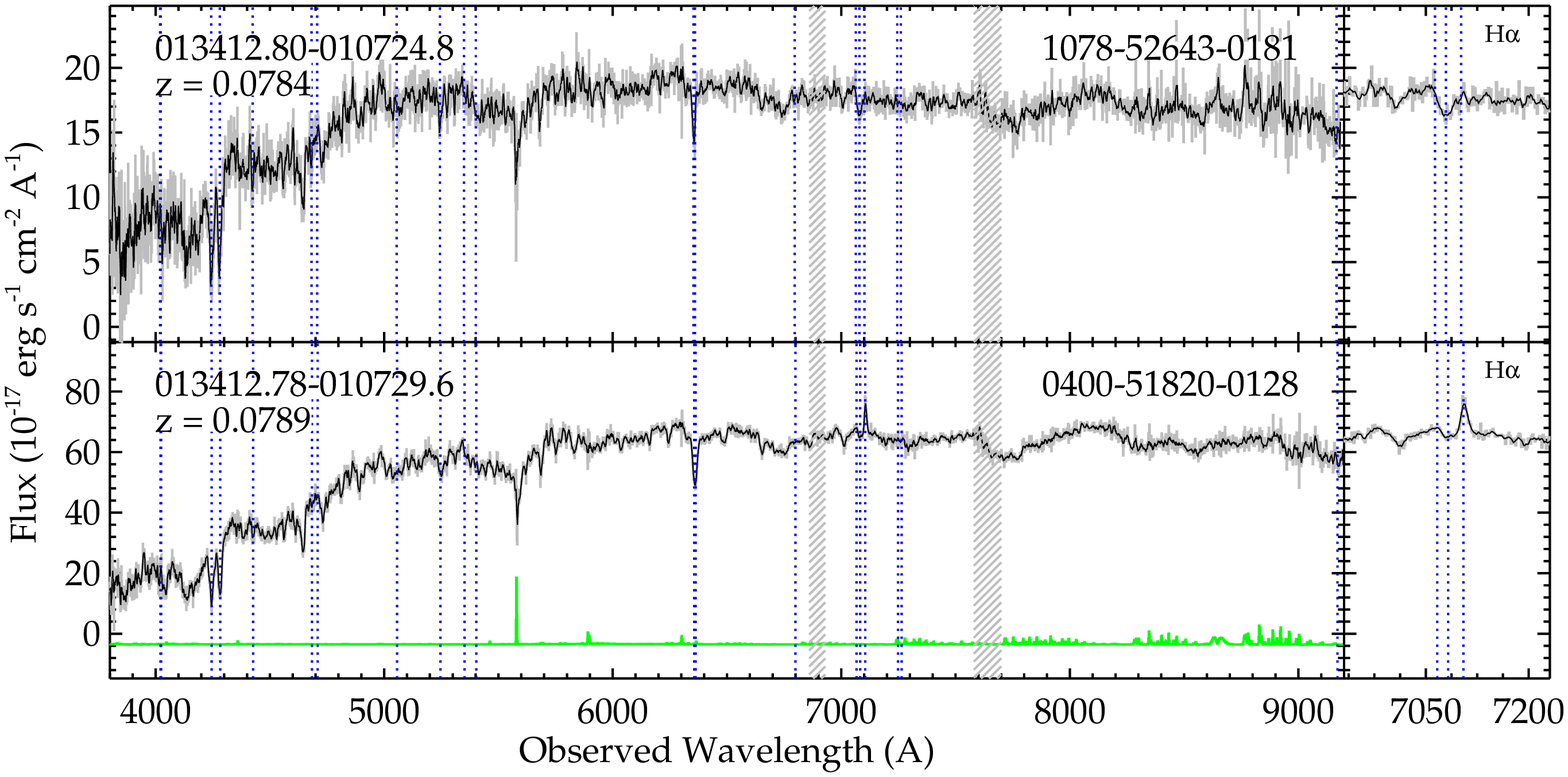}
\plottwo{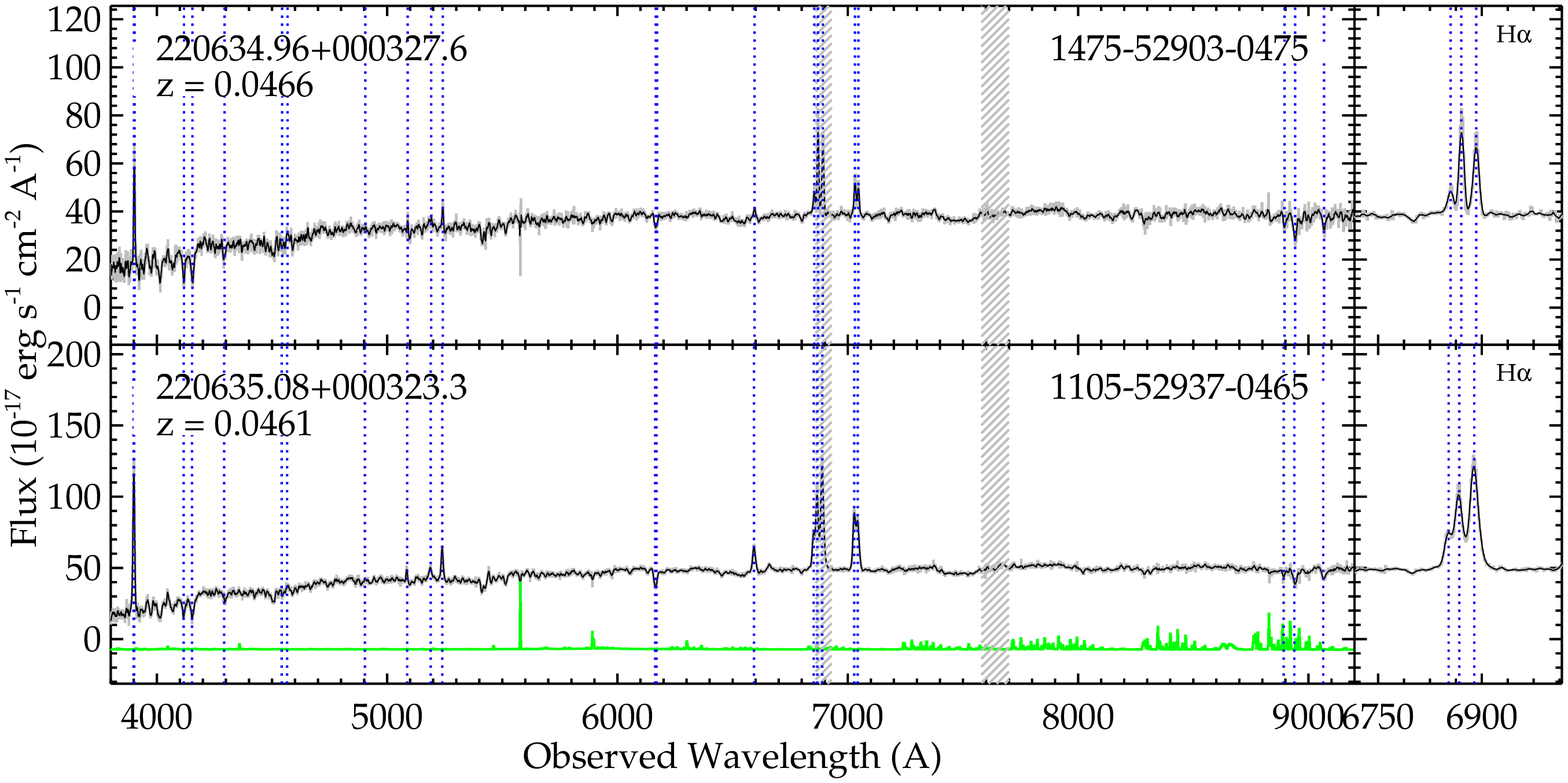}{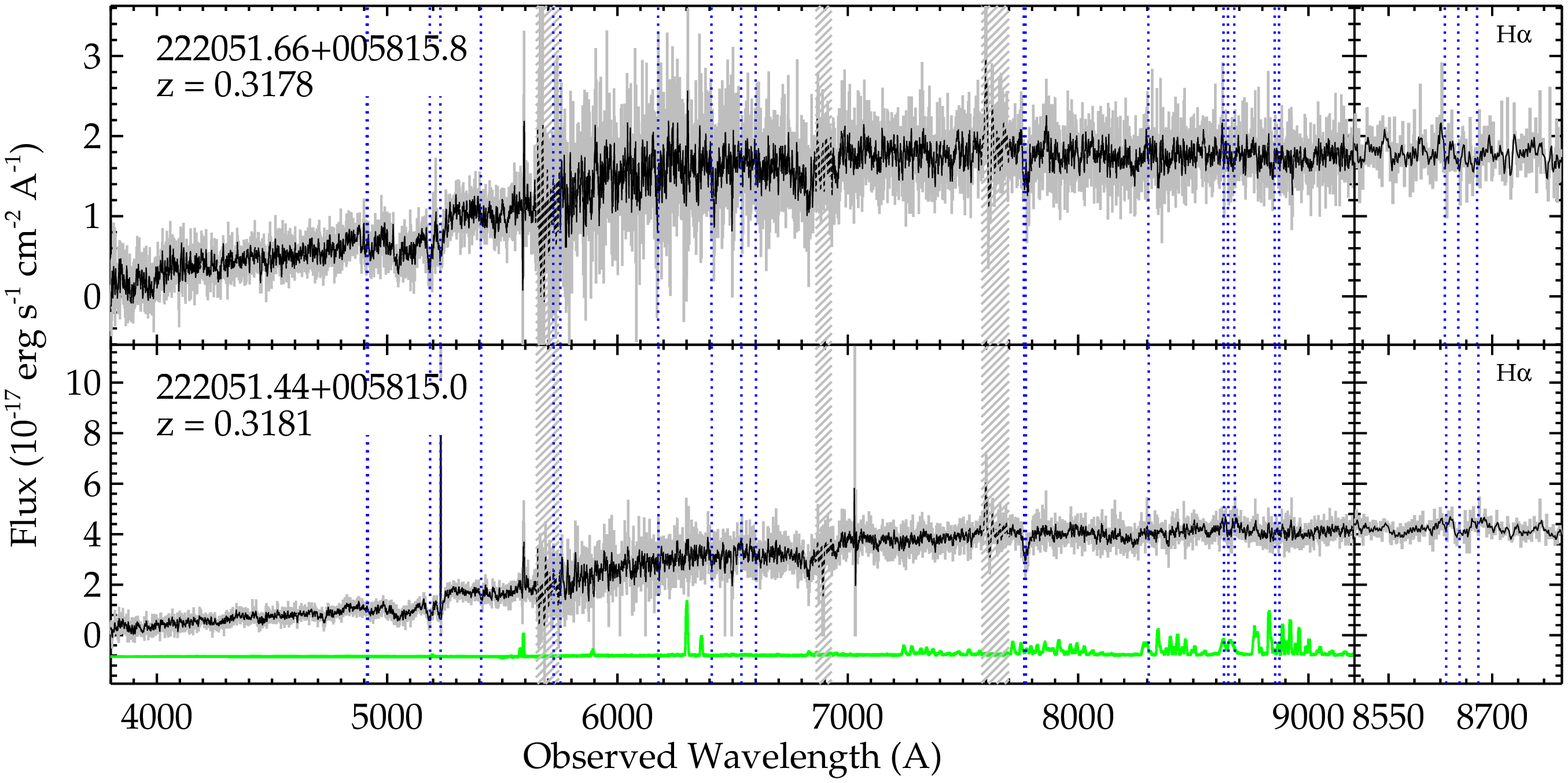}
\plottwo{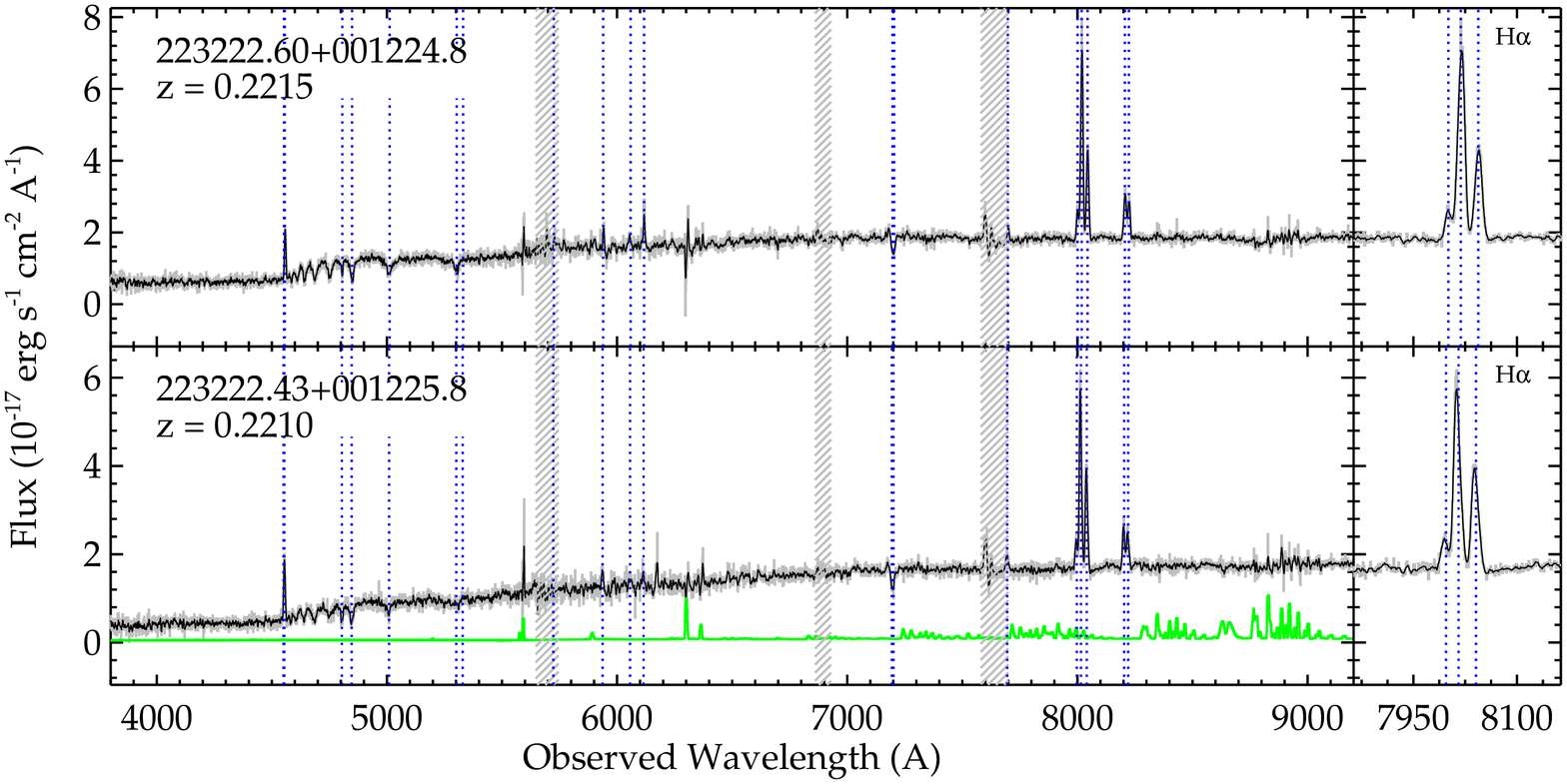}{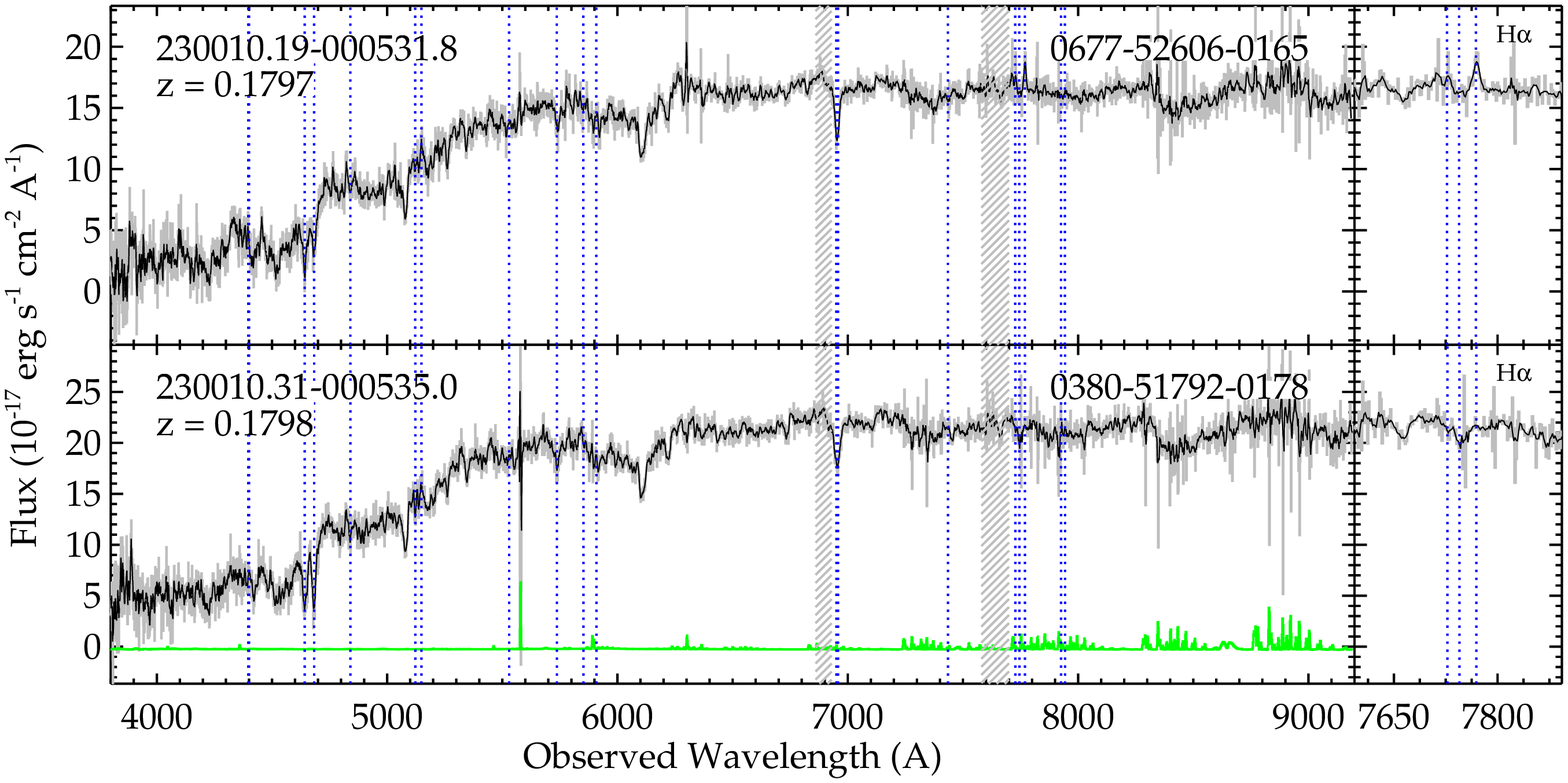}
\caption{Optical spectra for the candidate binaries that are spectroscopically confirmed mergers. Blue dotted vertical lines indicate the redshifted emission and absorption lines. Spectral regions affected by strong telluric absorption features and low LRIS dichroic transmission are hatched in gray. The smaller panels give a closer look at the H$\alpha$/[N\,{\sc ii}] region of each spectrum. A scaled sky spectrum is plotted in green at the bottom of each subfigure. For each spectrum, the corresponding radio source's designation is labeled, along with the spectroscopic redshift determined from stellar absorption features. The spectra for two pairs (2220$+$0058 and 2232$+$0012) are from Keck/LRIS, and the rest are from the SDSS. The SDSS spectra are labeled with the plate number, the Modified Julian Date (MJD), and the fiber ID on the top right. 
\label{fig:specs}}
\end{figure*}

Spectroscopy is useful to measure the redshifts and to determine the nature of the radio sources. Here we have obtained optical spectra for the radio sources from the SDSS and our own Keck program. Spectroscopic redshifts of the candidate binaries are included in Tables~\ref{tab:gradeA} and \ref{tab:gradeB}. We obtained spectroscopy for both components for eight candidate binaries (5 grade A's and 3 grade B's). Two out of the eight candidates turned out to be projected pairs (i.e., the optical IDs are at widely different redshifts; 2302$-$0003 and 2318$+$0045), and the other six are all confirmed mergers with radial velocity separations less than 140~\kms\ and projected physical separation between 4~kpc $<$ Sep $<$ 12~kpc. Properties of these confirmed pairs are listed in Table~\ref{tab:pairs}, and their spectra are shown in Figure~\ref{fig:specs}.

\subsection{SDSS Spectroscopy}

We cross-matched the radio sources' optical IDs with the SDSS DR10 spectroscopic catalog (SpecObj-dr10.fits). The DR10 catalog includes spectra from SDSS I, II and III, including the SDSS-I and II Legacy Surveys, SEGUE1, SEGUE2 and BOSS. We use a matching radius of 1.0\arcsec, $\sim2\times$ smaller than that used in matching the radio catalog with the photometric catalogs. 2,622 radio sources have SDSS spectra. Although the spectroscopic redshifts range from $-0.004$ to $4$, 75\% of the sources are at $z < 0.6$ and the median redshift is $\tilde{z} = 0.37$. 

Because two fibers on the same plate can not be placed closer than 55\arcsec\ and 65\arcsec\ for SDSS and BOSS, respectively, pairs of galaxies with small angular separations would suffer from incompleteness due to fiber collisions \citep{Blanton03a,Dawson13}, unless they are in overlapping regions of different plates. So we expect only a small fraction of the pairs in our sample have SDSS spectroscopy for both components. Indeed, only 4 out of the 44 candidate binaries have SDSS spectra for both components (Fig.~\ref{fig:specs}). Among those, two binaries overlap with the binary AGN sample selected from spectroscopically identified AGNs within SDSS DR7 \citep[0051$+$0020 and 2206$+$0003;][]{Liu11}. The other two, 0134$-$0107 and 2300$-$0005, are not in their sample because their optical emission comes almost entirely from old stellar populations without detectable emission lines. As we will show, galaxies in the latter pairs are {\it bona fide} radio AGNs.

\subsection{Keck LRIS Spectroscopy}

We obtained longslit spectroscopy during part of 2011 Oct 23 (UT) with the Low Resolution Imaging Spectrometer \citep[LRIS;][]{Oke95} on the Keck I telescope. We used the 600/4000 Grism in the blue, the 400/8500 grating in the red, and the 560 Dichroic. The central wavelength of the grating was 8000~\AA, so that the blue and red spectra overlap between 5595 and 5810~\AA, and the full wavelength range is from 3100~\AA\ to 1~\um. The 1.5\arcsec-wide longslit was used throughout the night. With the chosen dispersive elements, the spectral resolutions in FWHM are $\sim$6~\AA\ and $\sim$10~\AA\ for the blue and red spectra, respectively. The integration times ranged from 10 min to 30 min. The seeing was highly variable: the CFHT/DIMM's measurements ranged from 0.3\arcsec\ to $>$2\arcsec\ with a mean around 1.5\arcsec. The spectrophotometric standard star BD$+$28~4211 was observed at the beginning of the night for flux calibration. 

We reduced the raw data with the XIDL code\footnote{http://www.ucolick.org/$\sim$xavier/IDL/} written and maintained by J.~X.~Prochaska, J.~Hennawi, and others. The XIDL/Low-Redux pipeline follows the standard data reduction steps and reduces the blue and red channels separately. It begins by subtracting a super bias from the raw CCD frames, tracing the slit profiles using flat fields, and deriving the 2D wavelength solution for each slit using the arcs. Then it flat-fields each slit and rejects cosmic-rays, identifies objects automatically in the slit, and builds the 2D bspline super sky model. No rectification is performed to avoid interpolations. Before and after subtracting the super sky model, we extract 1D spectra from each object using a top hat profile that follows the continuum trace of the object itself or that of the brightest source on the slit if the object is too faint. Finally, instrument flexure is removed by tweaking the wavelength array of the extracted 1D spectra using isolated sky lines, and a sensitivity function is derived from the extracted standard star spectra and is applied to the science spectra. 

Seven candidate binaries were selected for Keck spectroscopy. The LRIS sample included 2 grade A candidates (2232$+$0012 and 2300$-$0005), 4 grade B candidates (0152$-$0010, 2220$+$0058, 2302$-$0003, and 2318$+$0045), and 1 grade D pair (2252$+$0106, which is shown in the lower left panel in the grade D examples in Fig.~\ref{fig:gradeCD}). All pairs are spatially resolved except 0152$-$0010, for which we were only able to measure the spectroscopic redshift for the brighter component, 015253.79$-$001005.6. Six of 13 resolved sources covered by LRIS already have SDSS spectra, and they are consistent with the LRIS spectra with similar S/N; so we have obtained new optical spectra for 7 radio sources. 

The LRIS spectra of the grade D pair, 2252$+$0106, shows that it is a merging pair of a red elliptical and a tidally distorted star-forming galaxy at $z = 0.072$. We do not discuss it further because of its lowered grade.  

\section{Spectral Analysis}

\begin{figure}[!t]
\plotone{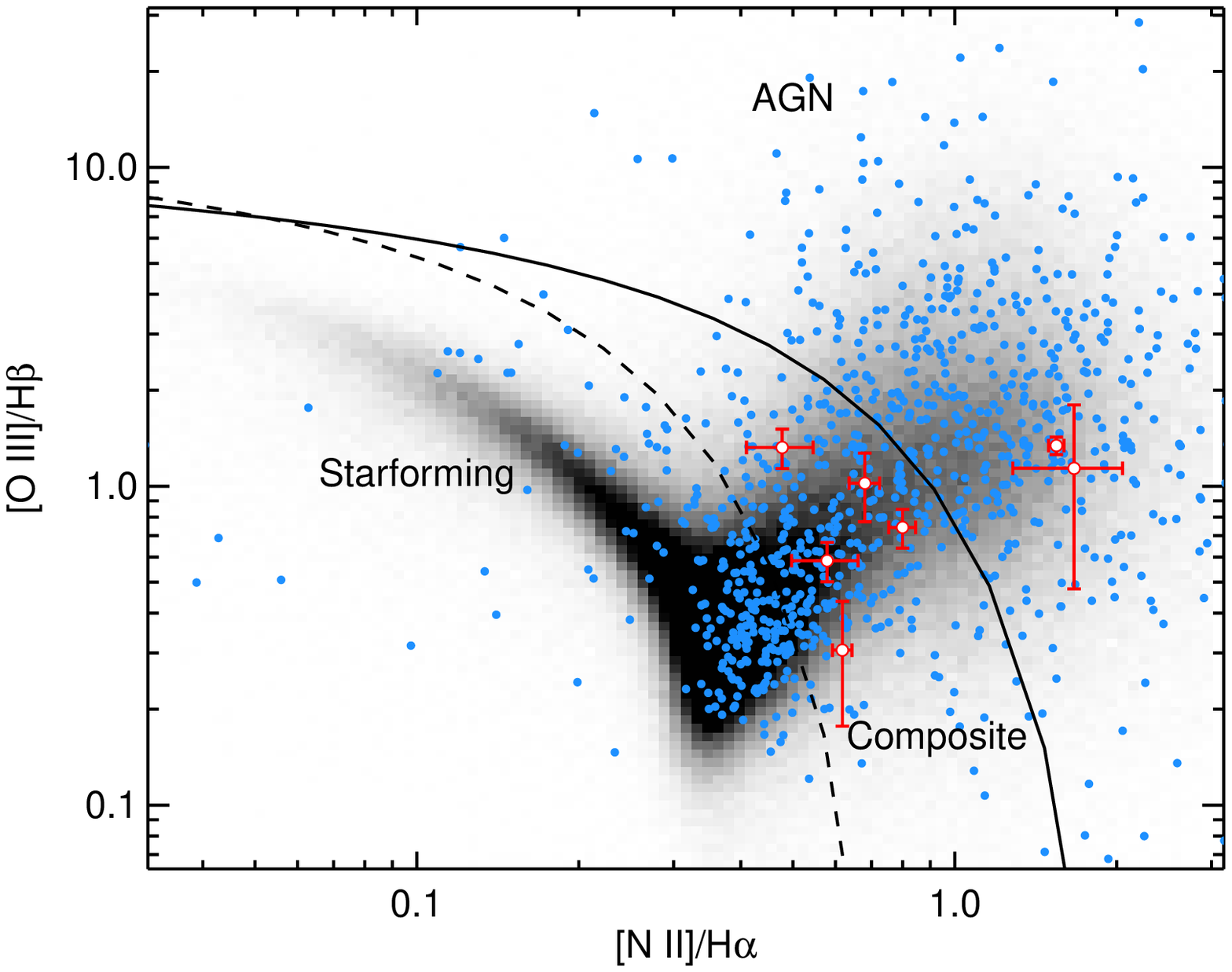}
\caption{Classical BPT diagnostic diagram of [O\,{\sc iii}]\,$\lambda$5007/H$\beta$ vs. [N\,{\sc ii}]~$\lambda$6583/H$\alpha$. The dashed line is the empirical star-forming line of \citet{Kauffmann03}, and the solid line is the theoretical maximum starburst line of \citet{Kewley01}. These demarcation lines separate the sources into starforming galaxies, AGN-starforming composites, and AGNs. The grey-scale background image shows the distribution of the SDSS DR8 emission-line galaxies. The blue points are the radio sources in the VLA-Stripe82 survey that show detectable optical emission lines. The red open circles with error bars show the 7 emission-line galaxies in the six spectroscopically confirmed grade A and B candidates (Table~\ref{tab:pairs}). \label{fig:BPT}}
\end{figure}

\begin{figure}[!t]
\plotone{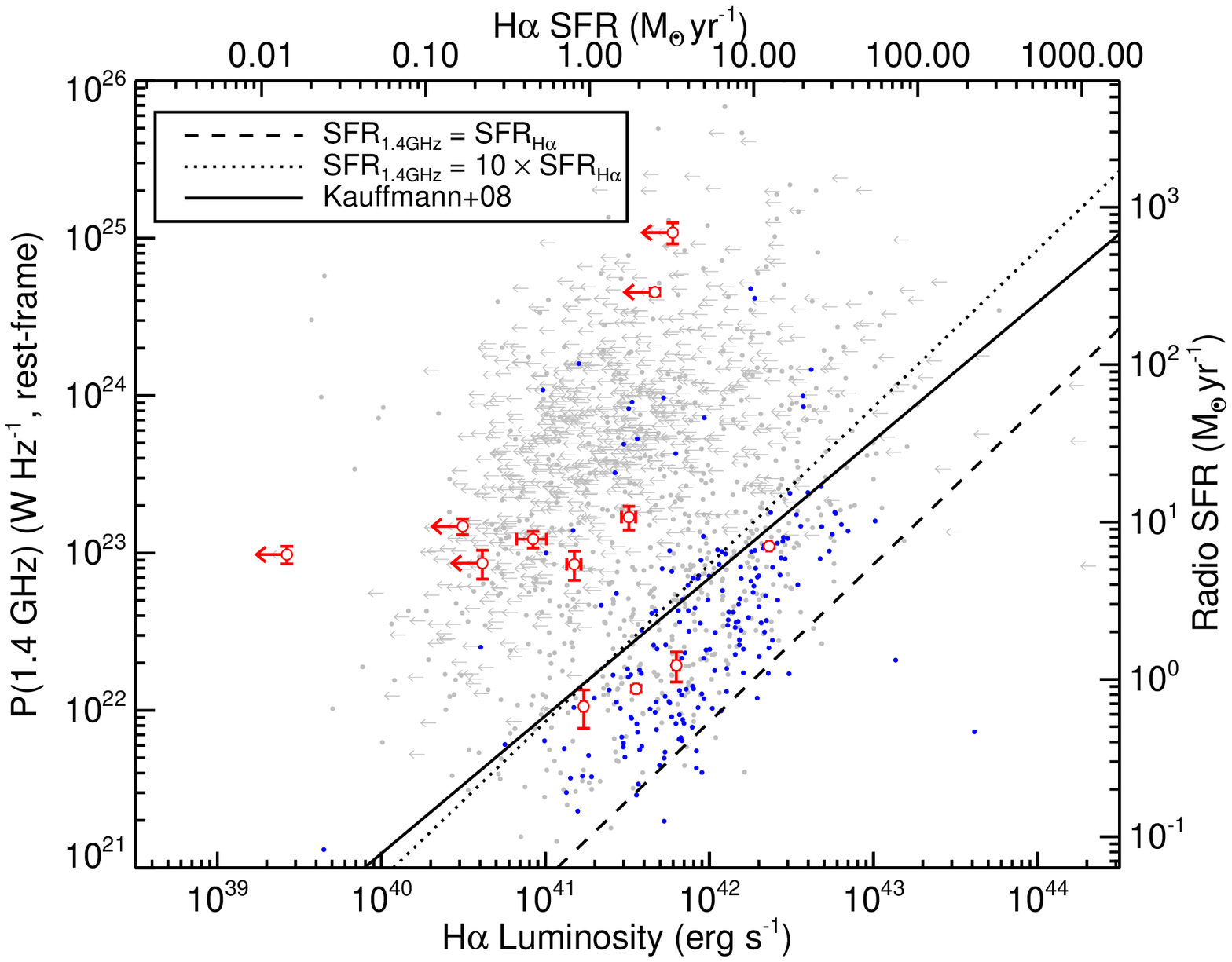}
\caption{Rest-frame 1.4~GHz radio power vs. H$\alpha$ luminosity. The H$\alpha$ luminosities have been corrected for dust-extinction and aperture loss. Top and right axes indicate the implied SFRs using the calibrations of \citet{Murphy13}. The solid line shows the empirical selection criterion for radio-excess AGNs \citep{Kauffmann08}, which is similar to requiring that the observed radio power is 10$\times$ beyond what is expected from the H$\alpha$-traced SFR. Gray points and arrows show the radio sources in the VLA-Stripe82 survey at $z < 0.4$. Starforming galaxies classified using the BPT diagram in Fig.~\ref{fig:BPT} are highlighted in blue. Most of these starforming galaxies lie below the dotted line, so we classify sources above the dotted line as radio-excess AGNs. Sources below the dotted line are radio-normal, and they could be either AGNs or starforming galaxies. The red open circles with error bars are the 12 galaxies in the six spectroscopically confirmed grade A and B candidates (Table~\ref{tab:pairs}). \label{fig:RP}}
\end{figure}

\subsection{The Radio-Selected Sample} \label{sec:radio_sample}

For the SDSS spectra, we adopt the emission line measurements from the Portsmouth group\footnote{https://www.sdss3.org/dr10/spectro/galaxy\_portsmouth.php} \citep{Thomas13}. The measurements are based on public codes Penalized Pixel-Fitting \citep[pPXF;][]{Cappellari04} and Gas AND Absorption Line Fitting \citep[GANDALF;][]{Sarzi06}. Stellar population templates and Gaussian emission lines are fit simultaneously to separate stellar continuum and emission/absorption lines from ionized gas. Dust-extinction correction is applied to both the stellar continuum and the emission lines. The same decomposition method is applied for the Keck/LRIS spectra. The emission line fluxes and their upper limits are corrected for fiber/slit aperture losses using the difference between $r$-band spectrum-synthesized magnitude and the \citet{Petrosian76} magnitude from the $r$-band image.

Out of the 2,629 radio sources with optical spectra, 1,386 are between $0 < z < 0.4$. We now discuss the properties of this subsample because we could measure H$\alpha$ fluxes, which is critical to classify the sources based on the \citet{Baldwin81} diagram (``BPT diagram'') involving the [N\,{\sc ii}]/H$\alpha$ ratio (Fig.~\ref{fig:BPT}) and the radio power vs. H$\alpha$ luminosity diagnostic diagram of \citet{Kauffmann08} (Fig.~\ref{fig:RP}). The BPT diagrams involve ratios of strong emission lines that are sensitive to ionizing spectra and gas metallicity but are insensitive to dust extinction. They have been widely used to distinguish between different ionization sources \citep[e.g.,][]{Kewley01,Kauffmann03,Kewley06}: e.g., star-forming regions, AGNs, and shocks or post-asymptotic giant branch stars \citep[e.g.,][]{Yan12}. 

After dust-extinction and aperture loss correction, the H$\alpha$ luminosity should be a good tracer of the total SFR. It thus can be used to predict the expected radio power from star formation alone \citep[e.g.,][]{Murphy13} because of the empirical IR-radio correlation of starforming galaxies \citep{Helou85,Yun01}. Any significant excess in the observed radio power would indicate AGN activity. As shown in Fig.~\ref{fig:RP}, most of the star-forming galaxies classified based on the BPT diagram (Fig.~\ref{fig:BPT}) have radio-derived SFRs less than 10$\times$ the H$\alpha$-derived SFRs. So we consider sources above the dotted line as radio-excess AGNs. Note that sources below the dotted line could still be AGNs, but they are radio-normal. 

We find that 82 of the 1,386 radio sources (6\%) are type-1 AGNs with broad emission lines (``AGN1''), 247 (18\%) are type-2 Seyferts/LINERs (``AGN2''), 209 (15\%) are AGN-starforming composite galaxies (``Comp''), 663 (48\%) are dominated by old or intermediate-age stellar populations with weak or undetectable emission lines (``Passive''), and only 185 (13\%) are starforming galaxies (``SF''). Excluding the passive sources, 26\% of the radio sources are starforming galaxies. In contrast, $\sim$50\% of the SDSS emission line galaxies are classified as starforming galaxies using the same BPT diagram. 

The AGN fraction of the radio-selected galaxies is extremely high. We find that 96\% of the Passive sources, which we cannot classify using the BPT diagram, are radio-excess AGNs, because their radio power is 10$\times$ greater than that expected from the level of star-forming activity traced by the H$\alpha$ luminosity (Fig.~\ref{fig:RP}). If we count AGN1, AGN2, Comps, and the radio-excess Passive sources all as AGNs, then the AGN fraction of our radio sample at $z < 0.4$ is $\sim$85\%.

The diverse spectroscopic properties and high AGN fraction of our radio-selected sample are consistent with previous results from the SDSS and FIRST surveys \citep[e.g.,][]{Best05,Kauffmann08a,Vitale12} and deep multi-wavelength extragalactic surveys \citep[e.g., AGES,][]{Hickox09}.

\subsection{Spectroscopically Confirmed Pairs}

Now we focus on the 6 Grade A and B candidates that are spectroscopically confirmed pairs (Table~\ref{tab:pairs}). These sources are plotted as open red circles with error bars in Figs~\ref{fig:BPT} and \ref{fig:RP}. Seven of the 12 galaxies in these pairs show detectable emission lines, and their line ratios indicate that they are either SF-AGN composites or AGN2s (Fig.~\ref{fig:BPT}). Except the source with the weakest emission lines (013412.78$-$010729.6), the observed H$\alpha$ equivalent widths ($W_{\rm H\alpha}$) are all significantly greater than 3~\AA, placing them safely above the ``retired galaxies'' on the WHAN diagram \citep{Cid-Fernandes11}. But as Fig.~\ref{fig:RP} shows, 013412.78$-$010729.6 is an radio-excess AGN because its H$\alpha$ luminosity is too low in comparison to its radio power. In Table~\ref{tab:pairs} we tabulate the emission line ratios of the seven sources along with their BPT classifications and H$\alpha$ equivalent widths.

The optical emission from the other five galaxies arises from old stellar populations, and the 3$\sigma$ luminosity upper limits of the undetected H$\alpha$ lines are insufficient to explain the high radio power, suggesting that they are radio-excess AGNs (Fig.~\ref{fig:RP}). Combining the above two AGN diagnostics, we find that all of the galaxies in the six confirmed mergers host active nuclei, making them compelling candidates for binary radio AGNs similar to SDSS~J150243.1+111557 \citep{Fu11b,Wrobel14}. It is not surprising to find that all of the confirmed pairs are binary AGNs, given the high AGN fraction ($\sim$85\%) of the radio-selected sample (see the previous subsection).

\section{Summary and Discussion}

We have developed a technique to identify kpc-scale binary AGNs with high-resolution wide-area radio surveys. Using the public VLA 1.4~GHz survey of $\sim$92~deg$^2$ of the SDSS Stripe 82 field, we identified 22 grade A and 30 grade B candidate binary AGNs. With SDSS and Keck spectroscopy, five out of five (100\%) grade A candidates and one out of three (33\%) grade B candidates are confirmed to be binary AGNs. Two of the three grade B candidates are projected pairs. The radio-detected galaxies in these candidates are classified as AGNs based on {\it either} excess in radio luminosity relative to H$\alpha$ luminosity {\it or} optical emission-line ratios that are diagnostics of the photo-ionizing source. We have started a follow-up campaign to obtain complete spectroscopic coverage for the remaining 44 candidates. Based on current results, we expect to confirm a total of $\sim$30 binary AGNs among the 52 candidates. Such a high success rate is expected thanks to the high AGN fraction of the radio-selected sample.

One of the main science questions to be addressed with a binary AGN sample is whether the AGN duty cycle is elevated during interactions. Our previous adaptive-optics and integral-field spectroscopy survey of a particular subsample of SDSS AGNs, the double-peaked [O\,{\sc iii}] AGNs, showed that the AGN duty cycle could be $\sim15\times$ higher in kpc-scale mergers than the average $\sim$1\% level. Because we confirmed 4 binary AGNs out of the 26 double-peaked AGNs that are also kpc-scale mergers. The quoted $\sim$1\% average duty cycle is for black holes of $M_{BH} \gtrsim 2\times10^6$~\msun\ at $z \sim 0.2$ with an Eddington ratio of $\lambda = L_{\rm bol}/L_{\rm Edd} \sim 0.4$ \citep{Shankar09}. In other words, the $\sim$1\% duty cycle is for AGNs with $L_{\rm bol} > 10^{44}$~\ergs\ at $z \sim 0.2$, consistent with the bolometric luminosities of the SDSS AGNs: their nuclear [O\,{\sc iii}]\,$\lambda$5007 line luminosities suggest bolometric luminosities of $L_{\rm bol} \simeq 3500 L_{\rm [O III]} \gtrsim 10^{44}$ \ergs\ \citep{Heckman04}. This result supports the notion that mergers could induce SMBH accretion by funneling gas deep into the nuclei. Furthermore, it is also consistent with recent simulations \citep[e.g.,][]{Van-Wassenhove12}, which showed that black hole accretion in a 2:1 gas-rich merger are not synchronized, instead, the simulated pair can be observed as a binary AGN only in 16\% of the time when either SMBH is an AGN (i.e., $L_{\rm bol} > 10^{44}$~\ergs) at a separation less than 10~kpc. 

Is the duty cycle of radio-mode accretion also elevated by galaxy interactions? We can make a rough estimate of the expected number of radio-loud binary AGNs by adopting the observed merger fraction of normal galaxies and the duty cycle of radio-loud AGNs. Given the $\sim$85\% AGN fraction of the spectroscopic subsample (\S~\ref{sec:radio_sample}), the entire VLA-Stripe82 catalog contains $\lesssim$15,000 AGNs. This is a strict upper limit because we have not removed extended sources that were split by the source-detection algorithm. Combining the local SMBH mass function of \citet{Shankar04} and the observed luminosity- and mass-dependent radio-loud fraction of \citet{Best05a}, we estimated a radio-loud duty cycle of $\sim$1\% for SMBHs more massive than $10^6$~\msun\ and above the luminosity limit of our AGN sample ($L_{\rm 1.4GHz} > 10^{22}$~W~Hz$^{-1}$). The lower limit of the SMBH mass corresponds to a stellar bulge mass of $7\times10^8$~\msun\ \citep{Haring04}, which is the mass limit of the SDSS survey at $z < 0.1$. Assuming that $\sim$5\% of the radio AGNs are in kpc-scale mergers \citep[as for normal galaxies;][]{Hopkins10b}, and adopting the radio-loud duty cycle of $\sim$1\%, we would expect to find at most $\sim$7 binaries. Suppose that we do confirm $\sim$30 binary AGNs, the simple estimate above shows that either the duty cycle or the merger fraction is underestimated by a factor of $\sim$4. Clearly, we need an independent determination of the merger fraction among radio-selected AGNs to break this degeneracy and obtain a reliable estimate of the duty cycle of radio-mode accretion in galaxy mergers.   

Our search of radio-selected binary AGNs complements binary searches in X-ray \citep[e.g.,][]{Koss12}, optical \citep[e.g.,][]{Comerford09a,Liu11}, or IR \citep[e.g.,][]{Tsai13}, because the majority of radio-selected AGNs show different properties from AGNs selected at other wavelengths \citep[e.g.,][]{Best07,Kauffmann08}. \citet{Hickox09} found that radio-selected AGNs at $z < 0.8$ primarily inhabit luminous red-sequence galaxies in massive ($\sim4\times10^{13}$~\msun) dark matter halos. Their accretion rates are lower than X-ray-selected and IR-selected AGNs, with Eddington ratios of $\lambda \lesssim 10^{-3}$. It has been suspected that the progenitors of radio-selected AGNs live in the most massive halos at $z \gtrsim 4$. Their host galaxies have gone through intense growth phases as merger-triggered starburst galaxies and quasars at $z \sim 2$. At the observed epoch ($z < 0.8$), star formation in the host galaxies has long ceased and the central black holes are presumably accreting hot halo gas in a radiatively inefficient mode \citep[e.g.,][]{Allen06}. Therefore, our radio search is sensitive to accreting black holes at a different evolutionary stage, in a different population of host galaxies, and in different galactic environments \citep[e.g.,][]{Best07}, when compared to X-ray/Optical/IR-selected binaries.

The major advantages of using high-resolution radio imaging to find binary AGNs are: (1) radio emission is free of dust obscuration, and (2) the relatively fast survey speed (at 1.4~GHz, each pointing of the VLA covers 0.25~deg$^2$ with an angular resolution of 1.4\arcsec).
Looking ahead, our method will be able to identify thousands of kpc-scale binary AGNs, as high-resolution radio surveys of even wider areas become available, e.g., the VLA 3~GHz survey of the entire 300~deg$^2$ Stripe 82 \citep{Mooley14}, the 10,000 deg$^2$ VLA Sky Survey (VLASS\footnote{https://science.nrao.edu/science/surveys/vlass}), and eventually an all sky survey with the Square Kilometer Array (SKA\footnote{https://www.skatelescope.org/key-documents/}). 

\acknowledgments

We thank the anonymous referee for their comments that helped improve the paper.
We thank J.~Hennawi and J.~X.~Prochaska for their help with the XIDL code, and Linghua Jiang for help with accessing their coadded images and catalogs. 
ADM was partially supported by NASA ADAP award NNX12AE38G and EPSCoR award NNX11AM18A and by NSF award 1211112.
SGD acknowledges a partial support from the NSF grants AST-1313422 and AST-1447922.
% Keck
The authors wish to recognize and acknowledge the very significant cultural role and reverence that the summit of Mauna Kea has always had within the indigenous Hawaiian community. We are most fortunate to have the opportunity to conduct observations from this mountain.

{\it Facilities}: VLA, Keck:I (LRIS), Sloan

%\clearpage

\begin{deluxetable*}{lcccccccccccc} 
\tablewidth{0pt}
\tablecaption{Grade A Candidate Binary Radio AGNs
\label{tab:gradeA}}
\tablehead{ 
\colhead{Radio Designation} & \colhead{$S^{\rm peak}_{\rm 1.4}$} & \colhead{$S^{\rm int}_{\rm 1.4}$} & \colhead{Optical ID}  & \colhead{Sep} & \colhead{$u$} & \colhead{$g$} & \colhead{$r$} & \colhead{$i$}& \colhead{$z$} & \colhead{$z_{\rm spec}$} & \colhead{Instru} \\ 
\colhead{J2000} & \colhead{mJy~bm$^{-1}$} & \colhead{mJy} & \colhead{J2000} & \colhead{\arcsec} & \colhead{AB} & \colhead{AB} & \colhead{AB} & \colhead{AB}& \colhead{AB} & \colhead{} & \colhead{} \\ 
\colhead{(1)} & \colhead{(2)} & \colhead{(3)} & \colhead{(4)} & \colhead{(5)} & \colhead{(6)} & \colhead{(7)} &  \colhead{(8)} & \colhead{(9)} & \colhead{(10)} & \colhead{(11)} & \colhead{(12)}
}
\startdata
004437.33$+$010132.5& 0.38$\pm$0.07& 0.96&004437.33$+$010132.5&4.6&\nod&24.1&23.4&24.9&\nod&\nod  &\NOD\\
004437.48$+$010136.4& 2.02$\pm$0.07& 3.99&004437.48$+$010136.5&4.6&\nod&\nod&\nod&23.0&22.2&\nod  &\NOD\\
005043.06$-$003045.2& 4.31$\pm$0.05&15.36&005043.07$-$003045.5&2.3&\nod&23.8&23.6&22.9&\nod&\nod  &\NOD\\
005043.18$-$003043.6& 3.09$\pm$0.05& 2.96&005043.16$-$003043.6&2.3&\nod&24.3&22.9&21.9&21.0&\nod  &\NOD\\
005113.93$+$002047.0& 0.34$\pm$0.05& 0.61&005113.93$+$002047.2&3.4&20.0&18.6&17.1&15.3&16.8&0.1124&SDSS\\
005114.10$+$002049.5& 2.50$\pm$0.05& 3.47&005114.11$+$002049.4&3.4&20.0&17.8&17.1&15.2&16.4&0.1126&SDSS\\
011156.46$-$000015.1&11.06$\pm$0.66&16.80&011156.44$-$000015.1&2.5&23.5&23.0&22.9&22.4&22.2&\nod  &\NOD\\
011156.58$-$000017.6&23.19$\pm$0.68&41.18&011156.52$-$000017.3&2.5&24.0&22.3&21.0&20.5&20.2&\nod  &\NOD\\
012839.28$+$011309.4& 1.41$\pm$0.09& 2.13&012839.29$+$011309.3&4.3&\nod&25.2&23.6&22.7&21.5&\nod  &\NOD\\
012839.42$+$011312.4& 0.90$\pm$0.09& 2.17&012839.44$+$011313.0&4.3&\nod&25.7&24.8&24.3&\nod&\nod  &\NOD\\
013412.78$-$010729.6& 5.47$\pm$0.54& 8.14&013412.78$-$010729.5&4.6&18.2&15.7&14.8&14.2&14.0&0.0789&SDSS\\
013412.80$-$010724.8& 5.00$\pm$0.56& 6.58&013412.84$-$010725.0&4.6&20.7&18.5&17.5&17.2&16.8&0.0784&SDSS\\
013638.30$-$002403.2& 0.36$\pm$0.06& 0.37&013638.29$-$002403.5&4.4&\nod&\nod&\nod&23.1&23.3&\nod  &\NOD\\
013638.53$-$002401.8& 0.36$\pm$0.06& 0.39&013638.56$-$002401.8&4.4&24.6&25.3&23.7&22.5&21.4&\nod  &\NOD\\
014443.79$+$000618.6& 0.31$\pm$0.06& 0.39&014443.82$+$000619.6&2.3&\nod&23.6&24.2&23.1&22.7&\nod  &\NOD\\
014443.87$+$000620.8& 0.43$\pm$0.06& 1.69&014443.90$+$000621.6&2.3&25.9&23.1&21.8&20.7&20.6&\nod  &\NOD\\
014715.82$-$000820.1& 0.63$\pm$0.09& 1.46&014715.86$-$000819.5&2.4&24.4&23.0&20.7&19.6&18.7&0.4727&BOSS\\
014715.96$-$000817.5& 0.45$\pm$0.09& 1.82&014715.94$-$000817.4&2.4&21.4&21.2&20.1&20.2&19.9&\nod  &\NOD\\
021020.94$-$005124.5& 0.51$\pm$0.07& 2.44&021021.00$-$005124.2&3.2&23.9&21.1&19.6&19.1&18.7&\nod  &\NOD\\
021021.16$-$005127.2& 0.53$\pm$0.07& 2.12&021021.13$-$005126.7&3.2&24.4&22.5&21.5&20.8&20.5&\nod  &\NOD\\
220634.96$+$000327.6& 0.51$\pm$0.07& 2.10&220634.97$+$000327.6&4.6&17.9&16.3&15.5&15.2&14.7&0.0466&SDSS\\
220635.08$+$000323.3& 1.75$\pm$0.07& 2.77&220635.08$+$000323.3&4.6&17.5&15.8&15.2&14.8&14.2&0.0461&SDSS\\
220758.28$+$010049.0& 2.05$\pm$0.06& 3.47&220758.26$+$010048.3&4.9&\nod&24.4&24.2&24.2&\nod&\nod  &\NOD\\
220758.46$+$010052.2&15.64$\pm$0.06&19.69&220758.46$+$010052.2&4.9&24.6&23.3&22.2&21.2&20.3&\nod  &\NOD\\
222548.38$+$003155.7& 2.64$\pm$0.07& 5.22&222548.37$+$003155.6&3.3&\nod&\nod&24.6&23.4&\nod&\nod  &\NOD\\
222548.39$+$003158.8& 0.44$\pm$0.07& 0.46&222548.40$+$003158.8&3.3&\nod&23.7&23.9&22.9&23.2&\nod  &\NOD\\
223222.43$+$001225.8& 0.37$\pm$0.06& 0.62&223222.41$+$001226.3&3.2&25.0&20.9&19.8&19.0&18.2&0.2210&LRIS\\
223222.60$+$001224.8& 0.53$\pm$0.06& 1.23&223222.60$+$001224.7&3.2&21.8&20.5&19.1&19.1&18.5&0.2215&LRIS\\
223546.30$+$000358.1& 0.40$\pm$0.06& 3.71&223546.28$+$000358.8&1.8&28.1&24.0&21.6&20.6&19.9&\nod  &\NOD\\
223546.45$+$000358.5& 0.49$\pm$0.06& 0.60&223546.40$+$000358.7&1.8&25.1&24.5&22.7&21.6&22.0&\nod  &\NOD\\
224204.26$+$003029.4& 0.49$\pm$0.06& 0.89&224204.25$+$003029.4&3.1&24.5&25.4&23.3&20.7&20.1&\nod  &\NOD\\
224204.45$+$003028.0& 0.44$\pm$0.06& 0.81&224204.45$+$003028.1&3.1&21.8&20.9&19.9&19.4&19.1&\nod  &\NOD\\
230010.19$-$000531.8& 0.71$\pm$0.05& 1.70&230010.18$-$000531.7&2.5&20.9&19.0&17.4&16.8&16.6&0.1797&SDSS\\
230010.31$-$000535.0& 0.40$\pm$0.05& 0.99&230010.24$-$000534.0&2.5&20.6&18.6&17.2&16.7&16.2&0.1798&SDSS\\
230342.76$-$011712.7& 0.96$\pm$0.14& 1.74&230342.74$-$011712.7&2.7&24.7&23.7&21.7&20.6&20.2&\nod  &\NOD\\
230342.91$-$011711.2& 0.73$\pm$0.13& 3.27&230342.91$-$011711.9&2.7&27.7&21.6&20.3&19.5&19.4&0.5676&BOSS\\
230453.08$-$010946.7& 2.27$\pm$0.06& 3.36&230453.04$-$010946.6&1.7&\nod&\nod&\nod&23.0&\nod&\nod  &\NOD\\
230453.19$-$010944.7& 2.80$\pm$0.06& 3.52&230453.13$-$010945.5&1.7&\nod&\nod&\nod&23.3&22.8&\nod  &\NOD\\
230559.14$+$002410.5& 0.52$\pm$0.05& 0.82&230559.17$+$002409.0&1.6&24.5&25.7&21.1&19.8&19.0&\nod  &\NOD\\
230559.18$+$002407.4& 0.76$\pm$0.05& 2.21&230559.19$+$002407.4&1.6&22.7&21.3&20.0&20.2&19.5&\nod  &\NOD\\
230858.73$+$002528.1& 0.51$\pm$0.09& 1.20&230858.68$+$002527.8&3.7&24.3&24.3&23.4&22.4&21.9&\nod  &\NOD\\
230858.76$+$002531.4& 2.37$\pm$0.09& 2.54&230858.75$+$002531.4&3.7&21.0&21.1&20.9&20.7&20.5&\nod  &\NOD\\
231539.23$-$011516.9& 4.80$\pm$0.19&10.79&231539.19$-$011516.7&2.9&\nod&\nod&24.1&22.3&21.9&\nod  &\NOD\\
231539.34$-$011518.5& 4.94$\pm$0.18& 9.41&231539.32$-$011518.9&2.9&\nod&24.3&24.0&\nod&\nod&\nod  &\NOD
\enddata
\tablecomments{
Every two lines is a pair.
Column 1: J2000 coordinate of the radio source; 
Column 2: 1.4~GHz peak flux density in mJy~bm$^{-1}$; 
Column 3: 1.4~GHz integrated flux density in mJy.
Column 4: J2000 coordinates of the identified optical counterpart;
Column 5: Angular separation between the optical counterparts in each pair;
Columns 6$-$10: SDSS $ugriz$-band Petrosian magnitudes in AB system;
Column 11: Spectroscopic redshift if available;
Column 12: Instrument used to obtain the spectroscopic redshift. 
}
\end{deluxetable*}

\begin{deluxetable*}{lcccccccccccc} 
\tablewidth{0pt}
\tablecaption{Grade B Candidate Binary Radio AGNs
\label{tab:gradeB}}
\tablehead{ 
\colhead{Radio Designation} & \colhead{$S^{\rm peak}_{\rm 1.4}$} & \colhead{$S^{\rm int}_{\rm 1.4}$} & \colhead{Optical ID}  & \colhead{Sep} & \colhead{$u$} & \colhead{$g$} & \colhead{$r$} & \colhead{$i$}& \colhead{$z$} & \colhead{$z_{\rm spec}$} & \colhead{Instru} \\ 
\colhead{J2000} & \colhead{mJy} & \colhead{mJy} & \colhead{J2000} & \colhead{\arcsec} & \colhead{AB} & \colhead{AB} & \colhead{AB} & \colhead{AB}& \colhead{AB} & \colhead{} & \colhead{} \\ 
\colhead{(1)} & \colhead{(2)} & \colhead{(3)} & \colhead{(4)} & \colhead{(5)} & \colhead{(6)} & \colhead{(7)} &  \colhead{(8)} & \colhead{(9)} & \colhead{(10)} & \colhead{(11)} & \colhead{(12)}
}
\startdata
004306.47$+$001422.3& 0.38$\pm$0.07& 5.42&004306.39$+$001423.1&4.8&\nod&25.7&\nod&\nod&\nod&\nod  &\NOD\\
004306.69$+$001424.5& 0.87$\pm$0.07& 2.63&004306.70$+$001424.5&4.8&24.2&21.3&19.8&19.3&19.1&\nod  &\NOD\\
005410.63$-$001220.5& 6.25$\pm$0.09&10.09&005410.70$-$001220.9&2.0&\nod&\nod&23.0&21.4&20.6&\nod  &\NOD\\
005410.81$-$001222.0& 4.85$\pm$0.09& 8.23&005410.78$-$001222.4&2.0&\nod&24.6&24.3&\nod&\nod&\nod  &\NOD\\
010413.64$-$004241.9& 0.53$\pm$0.05& 3.44&010413.62$-$004241.8&4.9&\nod&23.9&24.1&21.7&21.3&\nod  &\NOD\\
010413.77$-$004237.4& 1.01$\pm$0.05& 1.06&010413.75$-$004237.4&4.9&\nod&23.5&23.9&23.5&\nod&\nod  &\NOD\\
011613.73$+$005807.1& 3.02$\pm$0.06& 5.19&011613.73$+$005807.3&5.4&19.0&24.1&23.0&21.4&20.8&\nod  &\NOD\\
011613.77$+$005802.8& 0.78$\pm$0.06& 2.41&011613.77$+$005801.9&5.4&\nod&25.0&\nod&24.0&\nod&\nod  &\NOD\\
012050.39$-$004657.2& 0.42$\pm$0.07& 0.96&012050.37$-$004656.6&3.3&\nod&24.6&23.3&22.2&22.2&\nod  &\NOD\\
012050.63$-$004659.3& 0.96$\pm$0.07& 2.00&012050.55$-$004658.4&3.3&23.8&23.0&21.0&20.1&19.7&\nod  &\NOD\\
012825.45$-$005814.7& 0.75$\pm$0.05& 2.65&012825.43$-$005814.8&3.1&\nod&\nod&23.5&21.9&20.5&\nod  &\NOD\\
012825.71$-$005814.3& 0.55$\pm$0.05& 1.28&012825.61$-$005813.3&3.1&23.7&24.5&23.0&22.0&22.1&\nod  &\NOD\\
013328.06$+$000436.0& 5.52$\pm$0.74&43.44&013328.03$+$000436.5&3.7&24.7&24.8&23.3&21.0&20.5&\nod  &\NOD\\
013328.09$+$000440.3& 5.18$\pm$0.75& 5.18&013328.09$+$000440.1&3.7&\nod&25.3&23.6&22.4&21.4&\nod  &\NOD\\
013505.87$+$011910.9& 1.22$\pm$0.12& 2.09&013505.89$+$011911.5&3.2&21.0&20.3&18.4&17.7&17.4&0.3574&BOSS\\
013505.88$+$011908.9& 0.81$\pm$0.12& 0.86&013505.88$+$011908.4&3.2&22.7&22.2&22.0&21.5&20.9&\nod  &\NOD\\
013736.04$+$003636.1& 1.30$\pm$0.10& 2.42&013736.02$+$003637.8&4.5&\nod&\nod&25.5&\nod&\nod&\nod  &\NOD\\
013736.06$+$003633.3& 0.88$\pm$0.10& 0.94&013736.07$+$003633.3&4.5&23.3&24.1&22.1&20.8&20.1&\nod  &\NOD\\
014202.99$-$000150.1& 0.36$\pm$0.05& 0.44&014203.08$-$000150.3&2.5&24.7&24.3&22.5&21.1&21.1&\nod  &\NOD\\
014203.28$-$000151.0& 2.68$\pm$0.05& 4.07&014203.24$-$000151.0&2.5&22.2&24.6&21.9&20.5&20.0&\nod  &\NOD\\
014457.09$-$004216.7& 2.36$\pm$0.06& 3.43&014457.10$-$004216.6&2.5&24.8&24.8&22.4&20.3&19.5&\nod  &\NOD\\
014457.15$-$004218.0& 0.62$\pm$0.06& 5.39&014457.17$-$004218.8&2.5&22.0&20.9&19.7&19.3&18.9&\nod  &\NOD\\
014626.61$-$011436.7& 6.78$\pm$0.08& 9.89&014626.53$-$011436.4&3.7&\nod&\nod&24.2&\nod&\nod&\nod  &\NOD\\
014626.80$-$011439.5& 8.65$\pm$0.08&13.49&014626.72$-$011438.9&3.7&\nod&\nod&23.5&22.2&21.1&\nod  &\NOD\\
014655.08$-$001124.0& 0.67$\pm$0.06& 1.05&014655.06$-$001124.1&4.4&\nod&\nod&\nod&24.5&\nod&\nod  &\NOD\\
014655.31$-$001121.7& 2.18$\pm$0.06& 4.50&014655.32$-$001122.2&4.4&\nod&25.2&24.5&24.0&\nod&\nod  &\NOD\\
014928.27$-$001446.1& 0.55$\pm$0.05& 1.93&014928.39$-$001445.7&2.1&20.9&19.6&19.1&19.5&18.5&\nod  &\NOD\\
014928.50$-$001448.5& 0.45$\pm$0.05& 1.28&014928.41$-$001447.8&2.1&20.7&20.6&20.0&20.1&18.4&0.0897&BOSS\\
015253.79$-$001005.6& 0.70$\pm$0.06& 3.36&015253.79$-$001005.6&2.3&19.9&17.1&16.1&15.7&15.4&0.0824&SDSS\\
015253.91$-$001003.9& 0.33$\pm$0.06& 0.53&015253.92$-$001004.3&2.3&23.3&19.6&19.0&18.4&19.1&\nod  &\NOD\\
020301.36$-$003342.8& 0.49$\pm$0.10& 2.21&020301.40$-$003341.7&2.4&\nod&23.8&22.9&\nod&\nod&\nod  &\NOD\\
020301.42$-$003339.2& 1.00$\pm$0.10& 1.44&020301.43$-$003339.3&2.4&25.8&23.8&22.0&21.2&20.7&\nod  &\NOD\\
221029.52$-$000624.6& 0.41$\pm$0.07& 2.37&221029.41$-$000624.6&5.2&\nod&25.3&25.3&\nod&\nod&\nod  &\NOD\\
221029.64$-$000620.5& 3.55$\pm$0.07& 4.66&221029.62$-$000620.6&5.2&24.2&25.1&22.4&20.9&20.1&\nod  &\NOD\\
221142.39$+$002730.7& 2.64$\pm$0.07& 8.72&221142.42$+$002731.2&3.1&24.4&25.0&24.0&21.6&20.9&\nod  &\NOD\\
221142.61$+$002727.6& 5.39$\pm$0.07&12.73&221142.51$+$002728.5&3.1&20.1&19.8&19.6&19.3&19.1&0.7363&BOSS\\
221331.63$+$004835.1& 2.05$\pm$0.06& 4.90&221331.60$+$004834.6&4.3&\nod&\nod&23.9&25.5&23.6&\nod  &\NOD\\
221331.87$+$004836.1& 3.06$\pm$0.06& 5.39&221331.87$+$004836.1&4.3&21.0&24.8&21.6&20.6&19.8&\nod  &\NOD\\
221806.92$+$004515.6& 4.01$\pm$0.12& 4.79&221806.92$+$004515.9&2.2&\nod&\nod&\nod&23.0&\nod&\nod  &\NOD\\
221806.94$+$004517.5& 3.79$\pm$0.12& 5.24&221806.98$+$004517.9&2.2&22.1&22.9&22.3&21.8&21.2&\nod  &\NOD\\
222051.44$+$005815.0&12.27$\pm$0.57&14.72&222051.55$+$005815.5&2.6&23.4&20.4&18.5&17.9&17.5&0.3181&SDSS\\
222051.66$+$005815.8& 7.64$\pm$0.54&35.28&222051.70$+$005816.7&2.6&24.6&21.0&19.5&18.8&18.3&0.3178&LRIS\\
222907.42$-$000411.7& 3.82$\pm$0.06& 9.81&222907.53$-$000411.1&3.6&\nod&22.6&\nod&\nod&\nod&0.5935&BOSS\\
222907.73$-$000410.9& 3.24$\pm$0.06& 5.93&222907.77$-$000410.9&3.6&24.1&24.9&23.0&21.1&21.3&\nod  &\NOD\\
224426.38$+$001049.9& 0.86$\pm$0.08& 1.81&224426.44$+$001051.3&2.6&26.1&23.9&22.1&20.7&20.4&\nod  &\NOD\\
224426.55$+$001051.7& 0.71$\pm$0.08& 4.53&224426.61$+$001051.9&2.6&24.9&24.9&22.6&21.6&21.4&\nod  &\NOD\\
224532.53$+$005857.8& 1.36$\pm$0.06& 1.96&224532.53$+$005857.9&3.1&21.1&20.4&20.4&20.0&19.9&0.6492&SDSS\\
224532.71$+$005854.6& 0.49$\pm$0.06& 0.63&224532.68$+$005855.8&3.1&\nod&\nod&\nod&23.0&23.0&\nod  &\NOD\\
225002.03$+$001132.1& 0.38$\pm$0.06& 1.53&225002.04$+$001131.9&4.5&\nod&\nod&23.3&22.1&21.5&\nod  &\NOD\\
225002.10$+$001136.3& 0.30$\pm$0.06& 0.94&225002.10$+$001136.3&4.5&23.7&24.5&21.9&21.4&21.0&\nod  &\NOD\\
225817.73$+$003007.6& 1.05$\pm$0.09& 1.27&225817.73$+$003007.7&5.1&21.6&19.7&18.4&17.9&17.4&\nod  &\NOD\\
225817.91$+$003010.6& 0.85$\pm$0.09& 9.19&225817.97$+$003011.5&5.1&\nod&20.5&\nod&\nod&19.3&\nod  &\NOD\\
230223.16$-$000301.3& 0.34$\pm$0.06& 3.68&230223.28$-$000301.2&3.3&25.0&22.8&21.3&21.0&20.7&0.3110&LRIS\\
230223.46$-$000259.3& 0.45$\pm$0.06& 2.28&230223.46$-$000259.4&3.3&26.7&22.4&20.6&19.6&19.2&0.5455&BOSS\\
231405.96$+$002808.5& 0.39$\pm$0.06& 5.05&231406.04$+$002808.7&4.1&24.5&23.5&22.6&21.9&21.5&\nod  &\NOD\\
231406.05$+$002813.2& 2.45$\pm$0.06& 4.51&231406.03$+$002812.8&4.1&\nod&\nod&24.2&23.0&21.8&\nod  &\NOD\\
231843.31$+$004527.6& 0.35$\pm$0.06& 0.73&231843.30$+$004527.2&4.3&23.0&22.2&21.3&20.8&20.6&0.9600&LRIS\\
231843.56$+$004525.1& 1.84$\pm$0.06& 4.47&231843.55$+$004525.0&4.3&21.2&21.0&20.6&20.4&19.9&0.2750&LRIS\\
231953.40$+$003816.2& 2.26$\pm$0.06& 5.08&231953.31$+$003816.7&3.7&\nod&\nod&22.0&22.0&20.9&\nod  &\NOD\\
231953.40$+$003813.4& 0.94$\pm$0.06& 1.08&231953.43$+$003813.4&3.7&19.6&23.8&20.8&20.3&19.3&\nod  &\NOD
\enddata
\tablecomments{
Same as Table~\ref{tab:gradeA} but for grade B candidates.
}
\end{deluxetable*}

\begin{deluxetable*}{lcccccccccccccccc}
\small
\tablewidth{0pt}
\tablecaption{Properties of Spectroscopically Confirmed Pairs
\label{tab:pairs}}
\tablehead{ 
\colhead{Radio Designation} & \colhead{Grade} & \colhead{$z_{\rm spec}$} & \colhead{Instru} & \colhead{Sep} & \colhead{Sep} & \colhead{$\Delta V$} & \colhead{$P^{\rm int}_{\rm 1.4 GHz}$} & \colhead{$L_{\rm H\alpha}$} & \colhead{$W_{\rm H\alpha}$} & \colhead{[N\,{\sc ii}]/H$\alpha$} & \colhead{[O\,{\sc iii}]/H$\beta$} & \colhead{Radio} & \colhead{BPT} \\
\colhead{J2000} & \colhead{} & \colhead{} & \colhead{} & \colhead{\arcsec} & \colhead{kpc} & \colhead{\kms} &  \colhead{log(W~Hz$^{-1}$)} & \colhead{log(erg~s$^{-1}$)} & \colhead{\AA} & \colhead{} & \colhead{} & \colhead{Class} & \colhead{Class} \\
\colhead{(1)} & \colhead{(2)} & \colhead{(3)} & \colhead{(4)} & \colhead{(5)} & \colhead{(6)} & \colhead{(7)} &  \colhead{(8)} & \colhead{(9)} & \colhead{(10)} & \colhead{(11)} & \colhead{(12)} & \colhead{(13)} & \colhead{(14)} 
}
\startdata
005114.10$+$002049.5&A&0.1126&SDSS  &  3.4&  7.1& 58.7& 23.0&   42.4&51.0$\pm$0.3&0.6$\pm$0.1&0.3$\pm$0.1&Normal  &Comp\\
005113.93$+$002047.0&A&0.1124&SDSS  &  3.4&  7.1& 58.7& 22.3&   41.8&23.6$\pm$0.4&0.7$\pm$0.1&1.0$\pm$0.2&Normal  &Comp\\
013412.80$-$010724.8&A&0.0784&SDSS  &  4.6&  6.8&136.9& 23.0&$<$39.4&$-$3.0$\pm$0.9&       \nod&       \nod&Excess  &\nod\\
013412.78$-$010729.6&A&0.0789&SDSS  &  4.6&  6.8&136.9& 23.1&   40.9& 0.0$\pm$0.1&1.7$\pm$0.4&1.1$\pm$0.7&Excess  &AGN2\\
220634.96$+$000327.6&A&0.0466&SDSS  &  4.6&  4.2&124.5& 22.0&   41.2& 8.7$\pm$0.2&0.8$\pm$0.1&0.7$\pm$0.1&Normal  &Comp\\
220635.08$+$000323.3&A&0.0461&SDSS  &  4.6&  4.1&124.5& 22.1&   41.6&14.8$\pm$0.2&1.5$\pm$0.1&1.3$\pm$0.1&Normal  &AGN2\\
222051.66$+$005815.8&B&0.3178&LRIS  &  2.6& 11.9& 63.5& 25.0&$<$41.8& 0.0$\pm$0.4&       \nod&       \nod&Excess  &\nod\\
222051.44$+$005815.0&B&0.3181&SDSS  &  2.6& 11.9& 63.5& 24.7&$<$41.7&$-$2.0$\pm$1.2&       \nod&       \nod&Excess  &\nod\\
223222.60$+$001224.8&A&0.2215&LRIS  &  3.2& 11.6&122.8& 23.2&   41.5&35.6$\pm$0.3&0.5$\pm$0.1&1.3$\pm$0.2&Excess  &Comp\\
223222.43$+$001225.8&A&0.2210&LRIS  &  3.2& 11.6&122.8& 22.9&   41.2&28.3$\pm$0.3&0.6$\pm$0.1&0.6$\pm$0.1&Excess  &Comp\\
230010.19$-$000531.8&A&0.1797&SDSS  &  2.5&  7.7& 20.9& 23.2&$<$40.5&$-$0.1$\pm$0.2&       \nod&       \nod&Excess  &\nod\\
230010.31$-$000535.0&A&0.1798&SDSS  &  2.5&  7.7& 20.9& 22.9&$<$40.6&$-$1.5$\pm$0.3&       \nod&       \nod&Excess  &\nod
\enddata
\tablecomments{
Every two lines is a pair.
Column 1: J2000 coordinate of the radio source;
Column 2: grade of the candidate binary;
Column 3: spectroscopic redshift;
Column 4: instrument used to obtain the spectroscopic redshift;
Column 5: angular separation in arcsec between the optical counterparts in each pair;
Column 6: projected separation in kpc;
Column 7: radial velocity separation in \kms; 
Column 8: logarithmic radio power in W~Hz$^{-1}$ computed from the integrated source flux density
Column 9: logarithmic H$\alpha$ luminosity in erg~s$^{-1}$, corrected for reddening and aperture-loss;
Column 10: H$\alpha$ equivalent width in the observed frame in \AA;
Column 11: [N\,{\sc ii}]$\lambda$6584/H$\alpha$ line flux ratio and 1$\sigma$ error;
Column 12: [O\,{\sc iii}]$\lambda$5007/H$\beta$ line flux ratio and 1$\sigma$ error;
Column 13: radio classification based on the \citet{Kauffmann03} classification in Fig.~\ref{fig:RP};
Column 14: classification based on the BPT line-ratio diagram in Fig.~\ref{fig:BPT} (``Comp'' - AGN-starforming composite galaxy, ``AGN2'' - type-2 Seyfert or LINER).
}
\end{deluxetable*}


\begin{thebibliography}{87}
\expandafter\ifx\csname natexlab\endcsname\relax\def\natexlab#1{#1}\fi

\bibitem[{{Ahn} {et~al.}(2014){Ahn}, {Alexandroff}, {Allende Prieto}, {Anders},
  {Anderson}, {Anderton}, {Andrews}, {Aubourg}, {Bailey}, {Bastien}, \&
  et~al.}]{Ahn14}
{Ahn}, C.~P., {Alexandroff}, R., {Allende Prieto}, C., {et~al.} 2014, \apjs,
  211, 17

\bibitem[{{Allen} {et~al.}(2006){Allen}, {Dunn}, {Fabian}, {Taylor}, \&
  {Reynolds}}]{Allen06}
{Allen}, S.~W., {Dunn}, R.~J.~H., {Fabian}, A.~C., {Taylor}, G.~B., \&
  {Reynolds}, C.~S. 2006, \mnras, 372, 21

\bibitem[{{Annis} {et~al.}(2011){Annis}, {Soares-Santos}, {Strauss}, {Becker},
  {Dodelson}, {Fan}, {Gunn}, {Hao}, {Ivezic}, {Jester}, {Jiang}, {Johnston},
  {Kubo}, {Lampeitl}, {Lin}, {Lupton}, {Miknaitis}, {Seo}, {Simet}, \&
  {Yanny}}]{Annis11}
{Annis}, J., {Soares-Santos}, M., {Strauss}, M.~A., {et~al.} 2011,
  arXiv:1111.6619

\bibitem[{{Baker} {et~al.}(2008){Baker}, {Boggs}, {Centrella}, {Kelly},
  {McWilliams}, {Miller}, \& {van Meter}}]{Baker08}
{Baker}, J.~G., {Boggs}, W.~D., {Centrella}, J., {et~al.} 2008, \apjl, 682, L29

\bibitem[{Baldwin {et~al.}(1981)Baldwin, Phillips, \& Terlevich}]{Baldwin81}
Baldwin, J.~A., Phillips, M.~M., \& Terlevich, R. 1981, \pasp, 93, 5

\bibitem[{Becker {et~al.}(1995)Becker, White, \& Helfand}]{Becker95}
Becker, R.~H., White, R.~L., \& Helfand, D.~J. 1995, \apj, 450, 559

\bibitem[{{Beers} {et~al.}(1992){Beers}, {Gebhardt}, {Huchra}, {Forman},
  {Jones}, \& {Bothun}}]{Beers92}
{Beers}, T.~C., {Gebhardt}, K., {Huchra}, J.~P., {et~al.} 1992, \apj, 400, 410

\bibitem[{Begelman {et~al.}(1980)Begelman, Blandford, \& Rees}]{Begelman80}
Begelman, M.~C., Blandford, R.~D., \& Rees, M.~J. 1980, Nature, 287, 307

\bibitem[{{Berczik} {et~al.}(2006){Berczik}, {Merritt}, {Spurzem}, \&
  {Bischof}}]{Berczik06}
{Berczik}, P., {Merritt}, D., {Spurzem}, R., \& {Bischof}, H.-P. 2006, \apjl,
  642, L21

\bibitem[{{Berentzen} {et~al.}(2009){Berentzen}, {Preto}, {Berczik}, {Merritt},
  \& {Spurzem}}]{Berentzen09}
{Berentzen}, I., {Preto}, M., {Berczik}, P., {Merritt}, D., \& {Spurzem}, R.
  2009, \apj, 695, 455

\bibitem[{Bertin \& Arnouts(1996)}]{Bertin96}
Bertin, E., \& Arnouts, S. 1996, \aaps, 117, 393

\bibitem[{{Best} {et~al.}(2005){Best}, {Kauffmann}, {Heckman}, {Brinchmann},
  {Charlot}, {Ivezi{\'c}}, \& {White}}]{Best05a}
{Best}, P.~N., {Kauffmann}, G., {Heckman}, T.~M., {et~al.} 2005, \mnras, 362,
  25

\bibitem[{Best {et~al.}(2005)Best, Kauffmann, Heckman, \& Ivezic}]{Best05}
Best, P.~N., Kauffmann, G., Heckman, T.~M., \& Ivezic, Z. 2005, \mnras, 362, 9

\bibitem[{{Best} {et~al.}(2007){Best}, {von der Linden}, {Kauffmann},
  {Heckman}, \& {Kaiser}}]{Best07}
{Best}, P.~N., {von der Linden}, A., {Kauffmann}, G., {Heckman}, T.~M., \&
  {Kaiser}, C.~R. 2007, \mnras, 379, 894

\bibitem[{Bianchi {et~al.}(2008)Bianchi, Chiaberge, Piconcelli, Guainazzi, \&
  Matt}]{Bianchi08}
Bianchi, S., Chiaberge, M., Piconcelli, E., Guainazzi, M., \& Matt, G. 2008,
  \mnras, 386, 105

\bibitem[{{Blanton} {et~al.}(2003){Blanton}, {Lin}, {Lupton}, {Maley}, {Young},
  {Zehavi}, \& {Loveday}}]{Blanton03a}
{Blanton}, M.~R., {Lin}, H., {Lupton}, R.~H., {et~al.} 2003, \aj, 125, 2276

\bibitem[{{Blecha} {et~al.}(2013){Blecha}, {Civano}, {Elvis}, \&
  {Loeb}}]{Blecha13}
{Blecha}, L., {Civano}, F., {Elvis}, M., \& {Loeb}, A. 2013, \mnras, 428, 1341

\bibitem[{{Blecha}(2012)}]{Blecha12}
{Blecha}, L.~E. 2012, PhD thesis, Harvard University

\bibitem[{{Bonning} {et~al.}(2007){Bonning}, {Shields}, \&
  {Salviander}}]{Bonning07}
{Bonning}, E.~W., {Shields}, G.~A., \& {Salviander}, S. 2007, \apjl, 666, L13

\bibitem[{Cappellari \& Emsellem(2004)}]{Cappellari04}
Cappellari, M., \& Emsellem, E. 2004, \pasp, 116, 138

\bibitem[Cid Fernandes et al.(2011)]{Cid-Fernandes11} Cid Fernandes, 
R., Stasi{\'n}ska, G., Mateus, A., 
\& Vale Asari, N.\ 2011, \mnras, 413, 1687 

\bibitem[{{Colpi}(2014)}]{Colpi14}
{Colpi}, M. 2014, \ssr, 183, 189

\bibitem[{Comerford {et~al.}(2009)Comerford, Gerke, Newman, Davis, Yan, Cooper,
  Faber, Koo, Coil, Rosario, \& Dutton}]{Comerford09a}
Comerford, J.~M., Gerke, B.~F., Newman, J.~A., {et~al.} 2009, \apj, 698, 956

\bibitem[{{Dawson} {et~al.}(2013){Dawson}, {Schlegel}, {Ahn}, {Anderson},
  {Aubourg}, {Bailey}, {Barkhouser}, {Bautista}, {Beifiori}, {Berlind},
  {Bhardwaj}, {Bizyaev}, {Blake}, {Blanton}, {Blomqvist}, {Bolton}, {Borde},
  {Bovy}, {Brandt}, {Brewington}, {Brinkmann}, {Brown}, {Brownstein}, {Bundy},
  {Busca}, {Carithers}, {Carnero}, {Carr}, {Chen}, {Comparat}, {Connolly},
  {Cope}, {Croft}, {Cuesta}, {da Costa}, {Davenport}, {Delubac}, {de Putter},
  {Dhital}, {Ealet}, {Ebelke}, {Eisenstein}, {Escoffier}, {Fan}, {Filiz Ak},
  {Finley}, {Font-Ribera}, {G{\'e}nova-Santos}, {Gunn}, {Guo}, {Haggard},
  {Hall}, {Hamilton}, {Harris}, {Harris}, {Ho}, {Hogg}, {Holder}, {Honscheid},
  {Huehnerhoff}, {Jordan}, {Jordan}, {Kauffmann}, {Kazin}, {Kirkby}, {Klaene},
  {Kneib}, {Le Goff}, {Lee}, {Long}, {Loomis}, {Lundgren}, {Lupton}, {Maia},
  {Makler}, {Malanushenko}, {Malanushenko}, {Mandelbaum}, {Manera}, {Maraston},
  {Margala}, {Masters}, {McBride}, {McDonald}, {McGreer}, {McMahon}, {Mena},
  {Miralda-Escud{\'e}}, {Montero-Dorta}, {Montesano}, {Muna}, {Myers},
  {Naugle}, {Nichol}, {Noterdaeme}, {Nuza}, {Olmstead}, {Oravetz}, {Oravetz},
  {Owen}, {Padmanabhan}, {Palanque-Delabrouille}, {Pan}, {Parejko},
  {P{\^a}ris}, {Percival}, {P{\'e}rez-Fournon}, {P{\'e}rez-R{\`a}fols},
  {Petitjean}, {Pfaffenberger}, {Pforr}, {Pieri}, {Prada}, {Price-Whelan},
  {Raddick}, {Rebolo}, {Rich}, {Richards}, {Rockosi}, {Roe}, {Ross}, {Ross},
  {Rossi}, {Rubi{\~n}o-Martin}, {Samushia}, {S{\'a}nchez}, {Sayres}, {Schmidt},
  {Schneider}, {Sc{\'o}ccola}, {Seo}, {Shelden}, {Sheldon}, {Shen}, {Shu},
  {Slosar}, {Smee}, {Snedden}, {Stauffer}, {Steele}, {Strauss}, {Streblyanska},
  {Suzuki}, {Swanson}, {Tal}, {Tanaka}, {Thomas}, {Tinker}, {Tojeiro},
  {Tremonti}, {Vargas Maga{\~n}a}, {Verde}, {Viel}, {Wake}, {Watson}, {Weaver},
  {Weinberg}, {Weiner}, {West}, {White}, {Wood-Vasey}, {Yeche}, {Zehavi},
  {Zhao}, \& {Zheng}}]{Dawson13}
{Dawson}, K.~S., {Schlegel}, D.~J., {Ahn}, C.~P., {et~al.} 2013, \aj, 145, 10

\bibitem[{{Deane} {et~al.}(2014){Deane}, {Paragi}, {Jarvis}, {Coriat},
  {Bernardi}, {Fender}, {Frey}, {Heywood}, {Kl{\"o}ckner}, {Grainge}, \&
  {Rumsey}}]{Deane14}
{Deane}, R.~P., {Paragi}, Z., {Jarvis}, M.~J., {et~al.} 2014, \nat, 511, 57

\bibitem[{{Dotti} {et~al.}(2007){Dotti}, {Colpi}, {Haardt}, \&
  {Mayer}}]{Dotti07}
{Dotti}, M., {Colpi}, M., {Haardt}, F., \& {Mayer}, L. 2007, \mnras, 379, 956

\bibitem[{{Dotti} {et~al.}(2012){Dotti}, {Sesana}, \& {Decarli}}]{Dotti12}
{Dotti}, M., {Sesana}, A., \& {Decarli}, R. 2012, Advances in Astronomy, 2012

\bibitem[{Escala {et~al.}(2004)Escala, Larson, Coppi, \& Mardones}]{Escala04}
Escala, A., Larson, R.~B., Coppi, P.~S., \& Mardones, D. 2004, \apj, 607, 765

\bibitem[{Fabbiano {et~al.}(2011)Fabbiano, Wang, Elvis, \&
  Risaliti}]{Fabbiano11}
Fabbiano, G., Wang, J., Elvis, M., \& Risaliti, G. 2011, Nature, 477, 431

\bibitem[{Fu {et~al.}(2012)Fu, Yan, Myers, Stockton, Djorgovski, Aldering, \&
  Rich}]{Fu12a}
Fu, H., Yan, L., Myers, A.~D., {et~al.} 2012, \apj, 745, 67

\bibitem[{Fu {et~al.}(2011)Fu, Zhang, Assef, Stockton, Myers, Yan, Djorgovski,
  Wrobel, \& Riechers}]{Fu11b}
Fu, H., Zhang, Z.-Y., Assef, R.~J., {et~al.} 2011, \apjl, 740, L44

\bibitem[{Gerke {et~al.}(2007)Gerke, Newman, Lotz, Yan, Barmby, Coil,
  Conselice, Ivison, Lin, Koo, Nandra, Salim, Small, Weiner, Cooper, Davis,
  Faber, \& Guhathakurta}]{Gerke07}
Gerke, B.~F., Newman, J.~A., Lotz, J., {et~al.} 2007, \apj, 660, L23

\bibitem[{H{\"a}ring \& Rix(2004)}]{Haring04}
H{\"a}ring, N., \& Rix, H.-W. 2004, \apj, 604, L89

\bibitem[{Heckman {et~al.}(2004)Heckman, Kauffmann, Brinchmann, Charlot,
  Tremonti, \& White}]{Heckman04}
Heckman, T.~M., Kauffmann, G., Brinchmann, J., {et~al.} 2004, \apj, 613, 109

\bibitem[{Helou {et~al.}(1985)Helou, Soifer, \& Rowan-Robinson}]{Helou85}
Helou, G., Soifer, B.~T., \& Rowan-Robinson, M. 1985, \apj, 298, L7

\bibitem[{Hickox {et~al.}(2009)Hickox, Jones, Forman, Murray, Kochanek,
  Eisenstein, Jannuzi, Dey, Brown, Stern, Eisenhardt, Gorjian, Brodwin,
  Narayan, Cool, Kenter, Caldwell, \& Anderson}]{Hickox09}
Hickox, R.~C., Jones, C., Forman, W.~R., {et~al.} 2009, \apj, 696, 891

\bibitem[{{Hobbs} {et~al.}(2012){Hobbs}, {Coles}, {Manchester}, {Keith},
  {Shannon}, {Chen}, {Bailes}, {Bhat}, {Burke-Spolaor}, {Champion},
  {Chaudhary}, {Hotan}, {Khoo}, {Kocz}, {Levin}, {Oslowski}, {Preisig}, {Ravi},
  {Reynolds}, {Sarkissian}, {van Straten}, {Verbiest}, {Yardley}, \&
  {You}}]{Hobbs12}
{Hobbs}, G., {Coles}, W., {Manchester}, R.~N., {et~al.} 2012, \mnras, 427, 2780

\bibitem[{Hodge {et~al.}(2011)Hodge, Becker, White, Richards, \&
  Zeimann}]{Hodge11}
Hodge, J.~A., Becker, R.~H., White, R.~L., Richards, G.~T., \& Zeimann, G.~R.
  2011, \aj, 142, 3

\bibitem[{Hopkins {et~al.}(2010)Hopkins, Bundy, Croton, Hernquist, Keres,
  Khochfar, Stewart, Wetzel, \& Younger}]{Hopkins10b}
Hopkins, P.~F., Bundy, K., Croton, D., {et~al.} 2010, \apj, 715, 202

\bibitem[{{Hudson} {et~al.}(2006){Hudson}, {Reiprich}, {Clarke}, \&
  {Sarazin}}]{Hudson06}
{Hudson}, D.~S., {Reiprich}, T.~H., {Clarke}, T.~E., \& {Sarazin}, C.~L. 2006,
  \aap, 453, 433

\bibitem[{{Jiang} {et~al.}(2014){Jiang}, {Fan}, {Bian}, {McGreer}, {Strauss},
  {Annis}, {Buck}, {Green}, {Hodge}, {Myers}, {Rafiee}, \&
  {Richards}}]{Jiang14}
{Jiang}, L., {Fan}, X., {Bian}, F., {et~al.} 2014, \apjs, 213, 12

\bibitem[{Junkkarinen {et~al.}(2001)Junkkarinen, Shields, Beaver, Burbidge,
  Cohen, Hamann, \& Lyons}]{Junkkarinen01}
Junkkarinen, V., Shields, G.~A., Beaver, E.~A., {et~al.} 2001, \apjl, 549, L155

\bibitem[{{Kauffmann} {et~al.}(2008){Kauffmann}, {Heckman}, \&
  {Best}}]{Kauffmann08a}
{Kauffmann}, G., {Heckman}, T.~M., \& {Best}, P.~N. 2008, \mnras, 384, 953

\bibitem[{Kauffmann {et~al.}(2003)Kauffmann, Heckman, Tremonti, Brinchmann,
  Charlot, White, Ridgway, Brinkmann, Fukugita, Hall, Ivezic, Richards, \&
  Schneider}]{Kauffmann03}
Kauffmann, G., Heckman, T.~M., Tremonti, C., {et~al.} 2003, \mnras, 346, 1055

\bibitem[{Kauffmann {et~al.}(2008)Kauffmann, Bertoldi, Bourke, Evans, \&
  Lee}]{Kauffmann08}
Kauffmann, J., Bertoldi, F., Bourke, T.~L., Evans, N.~J., I., \& Lee, C.~W.
  2008, \aap, 487, 993

\bibitem[{Kewley {et~al.}(2001)Kewley, Dopita, Sutherland, Heisler, \&
  Trevena}]{Kewley01}
Kewley, L.~J., Dopita, M.~A., Sutherland, R.~S., Heisler, C.~A., \& Trevena, J.
  2001, \apj, 556, 121

\bibitem[{Kewley {et~al.}(2006)Kewley, Groves, Kauffmann, \&
  Heckman}]{Kewley06}
Kewley, L.~J., Groves, B., Kauffmann, G., \& Heckman, T. 2006, \mnras, 372, 961

\bibitem[{Komatsu {et~al.}(2011)Komatsu, Smith, Dunkley, Bennett, Gold,
  Hinshaw, Jarosik, Larson, Nolta, Page, Spergel, Halpern, Hill, Kogut, Limon,
  Meyer, Odegard, Tucker, Weiland, Wollack, \& Wright}]{Komatsu11}
Komatsu, E., Smith, K.~M., Dunkley, J., {et~al.} 2011, \apjs, 192, 18

\bibitem[{Komossa {et~al.}(2003)Komossa, Burwitz, Hasinger, Predehl, Kaastra,
  \& Ikebe}]{Komossa03}
Komossa, S., Burwitz, V., Hasinger, G., {et~al.} 2003, \apj, 582, L15

\bibitem[{Kormendy \& Richstone(1995)}]{Kormendy95}
Kormendy, J., \& Richstone, D. 1995, \araa, 33, 581

\bibitem[{Koss {et~al.}(2012)Koss, Mushotzky, Treister, Veilleux, Vasudevan, \&
  Trippe}]{Koss12}
Koss, M., Mushotzky, R., Treister, E., {et~al.} 2012, \apj, 746, L22

\bibitem[{Koss {et~al.}(2011)Koss, Mushotzky, Treister, Veilleux, Vasudevan,
  Miller, Sanders, Schawinski, \& Trippe}]{Koss11}
---. 2011, \apj, 735, L42

\bibitem[{Liu {et~al.}(2010)Liu, Shen, Strauss, \& Greene}]{Liu10a}
Liu, X., Shen, Y., Strauss, M.~A., \& Greene, J.~E. 2010, \apj, 708, 427

\bibitem[{{Liu} {et~al.}(2011){Liu}, {Shen}, {Strauss}, \& {Hao}}]{Liu11}
{Liu}, X., {Shen}, Y., {Strauss}, M.~A., \& {Hao}, L. 2011, \apj, 737, 101

\bibitem[{Madau \& Quataert(2004)}]{Madau04}
Madau, P., \& Quataert, E. 2004, \apj, 606, L17

\bibitem[{{Maness} {et~al.}(2004){Maness}, {Taylor}, {Zavala}, {Peck}, \&
  {Pollack}}]{Maness04}
{Maness}, H.~L., {Taylor}, G.~B., {Zavala}, R.~T., {Peck}, A.~B., \& {Pollack},
  L.~K. 2004, \apj, 602, 123

\bibitem[{Milosavljevi{\'c} \& Merritt(2001)}]{Milosavljevic01}
Milosavljevi{\'c}, M., \& Merritt, D. 2001, \apj, 563, 34

\bibitem[{Milosavljevic \& Merritt(2003)}]{Milosavljevic03}
Milosavljevic, M., \& Merritt, D. 2003, in The Astrophysics of Gravitational
  Wave Sources, Vol. 686, 201--210

\bibitem[{{Mooley} {et~al.}(2014){Mooley}, {Myers}, {Hallinan}, {Frail},
  {Kulkarni}, {Horesh}, \& {Bourke}}]{Mooley14}
{Mooley}, K.~P., {Myers}, S.~T., {Hallinan}, G., {et~al.} 2014, in American
  Astronomical Society Meeting Abstracts \#223, Vol. 223, \#236.02

\bibitem[{{Murphy} {et~al.}(2013){Murphy}, {Stierwalt}, {Armus}, {Condon}, \&
  {Evans}}]{Murphy13}
{Murphy}, E.~J., {Stierwalt}, S., {Armus}, L., {Condon}, J.~J., \& {Evans},
  A.~S. 2013, \apj, 768, 2

\bibitem[{{Newman} {et~al.}(2013){Newman}, {Cooper}, {Davis}, {Faber}, {Coil},
  {Guhathakurta}, {Koo}, {Phillips}, {Conroy}, {Dutton}, {Finkbeiner}, {Gerke},
  {Rosario}, {Weiner}, {Willmer}, {Yan}, {Harker}, {Kassin}, {Konidaris},
  {Lai}, {Madgwick}, {Noeske}, {Wirth}, {Connolly}, {Kaiser}, {Kirby},
  {Lemaux}, {Lin}, {Lotz}, {Luppino}, {Marinoni}, {Matthews}, {Metevier}, \&
  {Schiavon}}]{Newman13}
{Newman}, J.~A., {Cooper}, M.~C., {Davis}, M., {et~al.} 2013, \apjs, 208, 5

\bibitem[{Oke {et~al.}(1995)Oke, Cohen, Carr, Cromer, Dingizian, Harris,
  Labrecque, Lucinio, Schaal, Epps, \& Miller}]{Oke95}
Oke, J.~B., Cohen, J.~G., Carr, M., {et~al.} 1995, \pasp, 107, 375

\bibitem[{{Owen} {et~al.}(1985){Owen}, {O'Dea}, {Inoue}, \& {Eilek}}]{Owen85}
{Owen}, F.~N., {O'Dea}, C.~P., {Inoue}, M., \& {Eilek}, J.~A. 1985, \apjl, 294,
  L85

\bibitem[{{Peters}(1964)}]{Peters64}
{Peters}, P.~C. 1964, Physical Review, 136, 1224

\bibitem[{{Petrosian}(1976)}]{Petrosian76}
{Petrosian}, V. 1976, \apjl, 209, L1

\bibitem[{{Ravi} {et~al.}(2014){Ravi}, {Wyithe}, {Shannon}, {Hobbs}, \&
  {Manchester}}]{Ravi14}
{Ravi}, V., {Wyithe}, J.~S.~B., {Shannon}, R.~M., {Hobbs}, G., \& {Manchester},
  R.~N. 2014, \mnras, 442, 56

\bibitem[{Rodriguez {et~al.}(2006)Rodriguez, Taylor, Zavala, Peck, Pollack, \&
  Romani}]{Rodriguez06}
Rodriguez, C., Taylor, G.~B., Zavala, R.~T., {et~al.} 2006, \apj, 646, 49

\bibitem[{Sandage \& Matthews(1961)}]{Sandage61}
Sandage, A.~R., \& Matthews, T.~A. 1961, Sky \& Telescope, 21, 148

\bibitem[{Sarzi {et~al.}(2006)Sarzi, Falc{\'o}n-Barroso, Davies, Bacon, Bureau,
  Cappellari, de~Zeeuw, Emsellem, Fathi, Krajnovic, Kuntschner, McDermid, \&
  Peletier}]{Sarzi06}
Sarzi, M., Falc{\'o}n-Barroso, J., Davies, R.~L., {et~al.} 2006, \mnras, 366,
  1151

\bibitem[{{Scheel} {et~al.}(2009){Scheel}, {Boyle}, {Chu}, {Kidder},
  {Matthews}, \& {Pfeiffer}}]{Scheel09}
{Scheel}, M.~A., {Boyle}, M., {Chu}, T., {et~al.} 2009, \prd, 79, 024003

\bibitem[{Schmidt(1963)}]{Schmidt63}
Schmidt, M. 1963, Nature, 197, 1040

\bibitem[{{Schneider} {et~al.}(2010){Schneider}, {Richards}, {Hall}, {Strauss},
  {Anderson}, {Boroson}, {Ross}, {Shen}, {Brandt}, {Fan}, {Inada}, {Jester},
  {Knapp}, {Krawczyk}, {Thakar}, {Vanden Berk}, {Voges}, {Yanny}, {York},
  {Bahcall}, {Bizyaev}, {Blanton}, {Brewington}, {Brinkmann}, {Eisenstein},
  {Frieman}, {Fukugita}, {Gray}, {Gunn}, {Hibon}, {Ivezi{\'c}}, {Kent}, {Kron},
  {Lee}, {Lupton}, {Malanushenko}, {Malanushenko}, {Oravetz}, {Pan}, {Pier},
  {Price}, {Saxe}, {Schlegel}, {Simmons}, {Snedden}, {SubbaRao}, {Szalay}, \&
  {Weinberg}}]{Schneider10}
{Schneider}, D.~P., {Richards}, G.~T., {Hall}, P.~B., {et~al.} 2010, \aj, 139,
  2360

\bibitem[{{Sesana}(2013)}]{Sesana13}
{Sesana}, A. 2013, Classical and Quantum Gravity, 30, 244009

\bibitem[{Shankar {et~al.}(2004)Shankar, Salucci, Granato, De~Zotti, \&
  Danese}]{Shankar04}
Shankar, F., Salucci, P., Granato, G.~L., De~Zotti, G., \& Danese, L. 2004,
  \mnras, 354, 1020

\bibitem[{Shankar {et~al.}(2009)Shankar, Weinberg, \&
  Miralda-Escud{\'e}}]{Shankar09}
Shankar, F., Weinberg, D.~H., \& Miralda-Escud{\'e}, J. 2009, \apj, 690, 20

\bibitem[{Shen {et~al.}(2011)Shen, Liu, Greene, \& Strauss}]{Shen11}
Shen, Y., Liu, X., Greene, J.~E., \& Strauss, M.~A. 2011, \apj, 735, 48

\bibitem[{Smith {et~al.}(2010)Smith, Shields, Bonning, McMullen, Rosario, \&
  Salviander}]{Smith10}
Smith, K.~L., Shields, G.~A., Bonning, E.~W., {et~al.} 2010, \apj, 716, 866

\bibitem[{{Tanaka} \& {Haiman}(2013)}]{Tanaka13}
{Tanaka}, T.~L., \& {Haiman}, Z. 2013, Classical and Quantum Gravity, 30,
  224012

\bibitem[{{Thomas} {et~al.}(2013){Thomas}, {Steele}, {Maraston}, {Johansson},
  {Beifiori}, {Pforr}, {Str{\"o}mb{\"a}ck}, {Tremonti}, {Wake}, {Bizyaev},
  {Bolton}, {Brewington}, {Brownstein}, {Comparat}, {Kneib}, {Malanushenko},
  {Malanushenko}, {Oravetz}, {Pan}, {Parejko}, {Schneider}, {Shelden},
  {Simmons}, {Snedden}, {Tanaka}, {Weaver}, \& {Yan}}]{Thomas13}
{Thomas}, D., {Steele}, O., {Maraston}, C., {et~al.} 2013, \mnras, 431, 1383

\bibitem[{{Tsai} {et~al.}(2013){Tsai}, {Jarrett}, {Stern}, {Emonts}, {Barrows},
  {Assef}, {Norris}, {Eisenhardt}, {Lonsdale}, {Blain}, {Benford}, {Wu},
  {Stalder}, {Stubbs}, {High}, {Li}, \& {Kong}}]{Tsai13}
{Tsai}, C.-W., {Jarrett}, T.~H., {Stern}, D., {et~al.} 2013, \apj, 779, 41

\bibitem[{{Van Wassenhove} {et~al.}(2012){Van Wassenhove}, {Volonteri},
  {Mayer}, {Dotti}, {Bellovary}, \& {Callegari}}]{Van-Wassenhove12}
{Van Wassenhove}, S., {Volonteri}, M., {Mayer}, L., {et~al.} 2012, \apjl, 748,
  L7

\bibitem[{{Vitale} {et~al.}(2012){Vitale}, {Zuther},
  {Garc{\'{\i}}a-Mar{\'{\i}}n}, {Eckart}, {Bremer}, {Valencia-S.}, \&
  {Zensus}}]{Vitale12}
{Vitale}, M., {Zuther}, J., {Garc{\'{\i}}a-Mar{\'{\i}}n}, M., {et~al.} 2012,
  \aap, 546, A17

\bibitem[{Wang {et~al.}(2009)Wang, Chen, Hu, Mao, Zhang, \& Bian}]{Wang09}
Wang, J.-M., Chen, Y.-M., Hu, C., {et~al.} 2009, \apj, 705, L76

\bibitem[{{Wrobel} {et~al.}(2014){Wrobel}, {Walker}, \& {Fu}}]{Wrobel14}
{Wrobel}, J.~M., {Walker}, R.~C., \& {Fu}, H. 2014, \apjl, 792, L8

\bibitem[{{Yan} \& {Blanton}(2012)}]{Yan12}
{Yan}, R., \& {Blanton}, M.~R. 2012, \apj, 747, 61

\bibitem[{York {et~al.}(2000)York, Adelman, Anderson, Anderson, Annis, Bahcall,
  Bakken, Barkhouser, Bastian, Berman, Boroski, Bracker, Briegel, Briggs,
  Brinkmann, Brunner, Burles, Carey, Carr, Castander, Chen, Colestock,
  Connolly, Crocker, Csabai, Czarapata, Davis, Doi, Dombeck, Eisenstein,
  Ellman, Elms, Evans, Fan, Federwitz, Fiscelli, Friedman, Frieman, Fukugita,
  Gillespie, Gunn, Gurbani, de~Haas, Haldeman, Harris, Hayes, Heckman,
  Hennessy, Hindsley, Holm, Holmgren, Huang, Hull, Husby, Ichikawa, Ichikawa,
  Ivezi{\'c}, Kent, Kim, Kinney, Klaene, Kleinman, Kleinman, Knapp, Korienek,
  Kron, Kunszt, Lamb, Lee, Leger, Limmongkol, Lindenmeyer, Long, Loomis,
  Loveday, Lucinio, Lupton, MacKinnon, Mannery, Mantsch, Margon, McGehee,
  McKay, Meiksin, Merelli, Monet, Munn, Narayanan, Nash, Neilsen, Neswold,
  Newberg, Nichol, Nicinski, Nonino, Okada, Okamura, Ostriker, Owen, Pauls,
  Peoples, Peterson, Petravick, Pier, Pope, Pordes, Prosapio, Rechenmacher,
  Quinn, Richards, Richmond, Rivetta, Rockosi, Ruthmansdorfer, Sandford,
  Schlegel, Schneider, Sekiguchi, Sergey, Shimasaku, Siegmund, Smee, Smith,
  Snedden, Stone, Stoughton, Strauss, Stubbs, SubbaRao, Szalay, Szapudi,
  Szokoly, Thakar, Tremonti, Tucker, Uomoto, Vanden~Berk, Vogeley, Waddell,
  Wang, Watanabe, Weinberg, Yanny, \& Yasuda}]{York00}
York, D.~G., Adelman, J., Anderson, John~E., J., {et~al.} 2000, \aj, 120, 1579

\bibitem[{Yun {et~al.}(2001)Yun, Reddy, \& Condon}]{Yun01}
Yun, M.~S., Reddy, N.~A., \& Condon, J.~J. 2001, \apj, 554, 803

\bibitem[{Zakamska {et~al.}(2003)Zakamska, Strauss, Krolik, Collinge, Hall,
  Hao, Heckman, Ivezic, Richards, Schlegel, Schneider, Strateva, Vanden~Berk,
  Anderson, \& Brinkmann}]{Zakamska03}
Zakamska, N.~L., Strauss, M.~A., Krolik, J.~H., {et~al.} 2003, \aj, 126, 2125

\end{thebibliography}
\end{document}